\documentclass[useAMS,usenatbib]{mn2e}
\usepackage{epsfig}
\usepackage{amsmath}
\usepackage[authoryear]{natbib}
\usepackage{longtable}
\usepackage{graphicx}
\usepackage{amssymb}
\def\mh{\mbox{$M_{\rm h}$}} 
\def\mpc{h^{-1} {\rm{Mpc}}}
\def\kms {\rm{km~s^{-1}}}

\title[The nature of assembly bias - III]
  {The nature of assembly bias - III. 
Observational properties}
\author[Lacerna, Padilla $\&$ Stasyszyn]
  {Ivan Lacerna$^{1}$, Nelson Padilla$^{2, 3}$ and Federico Stasyszyn$^{4, 5}$ \\
    $^1$Instituto de Astronom\'ia, Universidad Nacional Aut\'onoma de M\'exico, A. P. 70-264, 04510, M\'exico, D.F., M\'exico \\
    $^2$Instituto de Astrof\'isica, Pontificia Universidad Cat\'olica de Chile, Av. V. Mackenna 4860, Santiago, Chile\\
    $^3$Centro de Astro-Ingenier\'ia, Pontificia Universidad Cat\'olica de Chile, 
Av. V. Mackenna 4860, Santiago, Chile\\
    $^4$Leibniz-Institut f\"ur Astrophysik Potsdam (AIP), An der Sternwarte
16, Postdam, Germany\\
    $^5$Universit\"ats Sternwarte M\"uenchen, Scheinerstr. 1, D-81679 M\"unchen, Germany 
}

\date{}

\pagerange{\pageref{firstpage}--\pageref{lastpage}} \pubyear{000}

\begin{document}

\label{firstpage}

\maketitle

%%%%%%%%%%%%%%%%%%%%%%%%%%%%%%%%%%%%%%%%%%%%%%%%%%%%%%%%%%%%%%%%%
\begin{abstract}
We analyse galaxies in groups
in the Sloan Digital Sky Survey (SDSS)  
and find a weak but significant assembly-type bias, 
where old central galaxies have a higher
clustering amplitude (61 $\pm$ 9 per cent) at scales $>$ 1 $\mpc$ than young central galaxies of equal host halo mass ($M_{h} \sim 10^{11.8} h^{-1}$ $M_{\odot}$). The observational sample is volume-limited out to $z=0.1$ with 
$M_r -$ 5 log$(h) \le -19.6$.
We construct a mock catalogue of galaxies that shows a similar signal of assembly bias (46 $\pm$ 9 per cent) at the same halo mass.
We then adapt the model presented by Lacerna \& Padilla (Paper I) to redefine the
overdensity peak height, 
which traces the assembly bias such that galaxies in equal
density peaks show the same clustering regardless of their stellar age,
but this time using observational features such as a flux limit. 
The proxy for peak height, which is proposed as a better alternative than the virial mass,
consists in the total mass given by the mass of neighbour host haloes in cylinders centred at each central galaxy.
The radius of the cylinder is parametrized as a function of stellar age and virial mass. The best-fitting sets of parameters 
that make the assembly bias signal lower than 5$-$15 per cent
for both SDSS and mock central galaxies are similar. 
The idea behind the parametrization is not to minimize the bias, but it is to use this method to understand the physical features that produce the assembly bias effect.
Even though the tracers of the density field used here differ significantly from those used in Paper I, our analysis of the simulated catalogue indicates that the different tracers produce correlated proxies, and therefore the reason behind this % 'this' added on 3 July 2014
assembly bias is the crowding of peaks in both simulations and the SDSS.
\end{abstract}
%%%%%%%%%%%%%%%%%%%%%%%%%%%%%%%%%%%%%%%%%%%%%%%%%%%%%%%%%%%%%%%%%
%%%%%%%%%%%%%%%%%%%%%%%%%%%%%%%%%%%%%%%%%%%%%%%%%%%%%%%%%%%%%%%%%
\section{Introduction}

In current hierarchical clustering models, galaxies form
by gas cooling inside virialized cold dark matter haloes %and it is assumed that
in a way that makes
their properties depend only on the halo mass in which they form. 
In the last few years,
results from $N$-body simulations have shown that the %, at a given mass,
halo clustering at large-scales depends on other halo %formation history
properties in
addition to the halo mass
\citep[e.g.][]{Gao05,Wechsler06,Zhu06,Gao-White07,FW10,Lacerna11,Lacerna12,vanDaalen12}.
%several halo properties are not independent of the surrounding
%larger-scale environment
%\cite{Gao05,Wechsler06,Bett07,Gao-White07,HT10,FW10}.
This effect in haloes, that is not expected from the
excursion set theory, was termed `assembly bias'.
For example, 
%\cite{Gao05}  %Gao et al. (2005) 
%found that 
low-mass haloes assembled at high redshifts are more strongly correlated than those
of the same mass assembled more recently.
Therefore, one could ask whether the assembly bias could %also 
affect the assumption that 
galaxy properties, e.g. galaxy clustering, depend only on the host halo mass.

By means of a semi-analytic model of galaxy formation, \cite{Croton07} %Croton et al. (2007) 
showed that galaxy clustering on large scales is affected by the correlation of 
halo environment with halo assembly history.
\citet[][hereafter Paper I]{Lacerna11} %Lacerna et al. (2011, hereafter Paper I) 
found an assembly bias effect in the 
semi-analytic galaxies from the \cite{LCP08} %Lagos, Cora $\&$ Padilla (2008) 
model,
where the old population has a higher clustering at large scales than the young population with the 
same host halo mass. 
Using semi-analytic models,
\cite{Xie14} found that the assembly bias may result in an abundance fraction 25 per cent lower for void galaxies than the expected by using their stellar-to-halo mass relation. %compared to field galaxies hosted by 
%haloes of equal mass.

On the observational side, the %environmental dependence is unclear.
existence of the assembly bias is still debated.
For example, 
\citet[][see also \citealt{Skibba_Sheth_2009}]{Skibba06}
used 
galaxies from the
Sloan Digital Sky Survey \cite[SDSS;][]{York00} %(SDSS; York et al. 2000) 
to measure marked correlation functions 
that when analysed using the halo model
suggest that 
the correlation between halo formation and environment is not important for the bright
galaxy population. 
\cite{Tinker11} %Tinker, Wetzel $\&$ Conroy (2011) 
found similar results, indicating that this dependence arises
just from the correlation between halo mass function and environment.
However, \cite{Y06} %Yang, Mo, \& van den Bosch (2006) 
and \cite{Wang08}
%Wang et al. (2008)
found that groups with red central galaxies, selected from the Two Degree Field Galaxy Redshift Survey
\citep[2dFGRS;][]{Colless01}
%(\mbox{2dFGRS}; Colless et al. 2001)
and SDSS respectively, 
are more strongly clustered than groups of the same mass but with blue central galaxies,
this effect being much more important for less massive groups.
%In addition to the clustering amplitude, 
\cite{Zapata09}
%Zapata et al. (2009) 
found that SDSS galaxy groups of similar mass and different
assembly histories show differences in their %galaxy population, for example in the
fraction of red galaxies.
Furthermore, \cite{Cooper10} %Cooper et al. (2010)
studied the relationship between the local environment
and %the 
properties of galaxies %which lie on
in the red sequence.
After removing the dependence of the average
overdensity on colour and stellar mass
by setting a density parameter as an estimator of environment %such as it is 
independent of
these galaxy properties,
they still found
%a strong dependence on the luminosity-weighted stellar age. 
that galaxies with older
stellar populations occupy regions of higher overdensities compared
to younger galaxies of similar colours or stellar masses.
Also, \cite{Alonso12} %Alonso et al. (2012) 
show that galaxies in groups in
low global densities are prone to show higher star formation %activity
compared to galaxies of equal properties and 
similar host groups but in
high density environments.
Recently, \cite{WangL13} have claimed the direct detection of the assembly bias in observations. They found that SDSS central galaxies with low specific star formation rates (sSFR) cluster more than those with %higher
high sSFR in low stellar mass bins. 
These results, from models and observations, show that the concept of assembly bias 
is applicable
to galaxies in addition to dark matter (DM) haloes
and would then 
influence the physics of galaxy formation.

The assembly bias effect is important when
galaxies are used to constrain cosmological parameters. 
\cite{Wu08} %Wu et al. (2008) 
showed that upcoming photometric catalogues such as the 
Dark Energy Survey (DES)\footnote{See http://www.darkenergysurvey.org/.} and the Large Synoptic Survey Telescope (LSST; LSST Science Collaborations et al. 2009)\footnote{See http://www.lsst.org/lsst/.} can
infer significantly biased cosmological parameters from the observed clustering
amplitude of galaxy clusters if the assembly bias is not taken into account.
Furthermore, the halo-galaxy connection inferred by models using galaxy clustering and assuming only halo mass dependence, such as the halo occupation distribution \citep[HOD,][]{BW02} and the conditional luminosity function \citep{Y03}, can be subject to an additional systematic error due to the assembly bias \citep[][]{WangDeLuciaWeinmann13,Zentner13}.
\citet{Hearin+2014} showed that an assembly bias effect in HOD models is required for  producing the signal of galactic conformity on Mpc scales, i.e. quenched central galaxies tend to reside preferentially in quenched large-scale environments \citep{Kauffmann+2013}. Therefore, some properties of central galaxies such as the quenching do not depend on the halo mass alone.

Paper I presented a new %model for 
peak height definition different from the halo virial mass,
%With this new definition 
where semi-analytic galaxies 
in haloes of a given mass %range
but different age 
do not show 
the assembly bias effect
in the two-halo regime.
This new definition simply adds mass to that of the dark matter halo
in a way that makes
old, low-mass haloes 
suffer
the most extreme peak height change
with respect to what results of using the virial mass as its proxy.
In this paper we will test
this model using the galaxy catalogue of 
\cite{Yang+2007}
based on 
SDSS. 
However,
instead of using the total
underlying mass inside spheres of different radii, we will use the total mass of haloes inside
cylinders along the position of each central galaxy
in order to make a more adequate proxy for a peak height, in a similar way
to what was proposed in Paper I
but applicable to real galaxy catalogues.
We will discuss the implications of using 
discrete 
tracers of the density field that consider
neighbour distinct haloes instead of a smooth density field.

The outline of this paper is as follows.
In Section \ref{sec_data} %Sec2%, 
we present a galaxy catalogue based on the SDSS and also
a 
mock catalogue of galaxies. 
In Section \ref{section_two-point_SDSS} we measure the assembly bias effect,
namely, the large-scale clustering of galaxies of a given host halo mass
depends on the age, for both the SDSS and the mock central galaxies.
We adopt a modified version of the model from Paper I %for the semi-analytic galaxies 
in Section \ref{sec_prm} using observational features.
The nature  of the objects %that are being 
considered
by using the proxy of the overdensity peak height is
shown in Section \ref{sec_peak_lum}, where we discuss the implications of using
discrete tracers of the density field that
consider neighbour haloes instead of a smooth density field.
%luminosity as a proxy of an overdensity peak height.
The conclusions of our results are presented in Section \ref{conclusiones}.

Throughout this paper we use the reduced Hubble constant $h$, where $H_0 = 100h$ km s$^{-1}$ Mpc$^{-1}$, with the following dependences: stellar age in $h^{-1}$ yr, halo mass in $h^{-1}$ $M_{\odot}$, scales (distances) in $h^{-1}$ Mpc, 
and absolute magnitudes in $+ 5$ log$(h)$,
unless the explicit value of $h$ is specified.

%%%%%%%%%%%%%%%%%%%%%%%%%%%%%%%%%%%%%%%%%%%%%%%%%%%%%%%%%%%%%%%%%
\section{Data}
\label{sec_data}

%%%sub-SECTION
\subsection{SDSS galaxies}
\label{section_SDSS_gal}

We use the galaxy group catalogue constructed by \citet[][hereafter Y07]{Yang+2007} 
from the 
New York University Value-Added Galaxy Catalog
%(NYU-VAGC, Blanton+2005),
\citep[NYU-VAGC;][]{Blanton+2005},
which is based on SDSS DR4 \citep[][]{DR4+2006}.
Y07 %\citep[see also][]{Yang+2008,Yang+2009,Yang+2012}
%(see also Yang+2008,Yang+2009,Yang+2012)
dynamically associated a dark matter halo to the galaxies
by using a halo-based group finder algorithm.
Each halo contains one or more galaxies out to its virial radius. 
The group finder consists of an iterative
procedure that uses average mass-to-light ratios of groups, based on the total stellar mass (or luminosity) of all group members
down to some luminosity, to assign a tentative mass to each group. Then the virial radius associated to this
mass is used for updating the group membership, repeating this process until convergence is reached.
%Y07 have tested this method with mock catalogs constructed based on the
%SDSS and they found that  $80\%$ have a completeness greater than 0.6,
%while $85\%$ have a contamination by interlopers lower than 0.5.
%We use the definition of the most massive galaxy 
%within a halo as the central galaxy.
The full sample consists of 369 447 galaxies with redshifts in the range 0.01 $\leq$ $z$ $\leq$ 0.2.
%We select the central galaxy as the most massive galaxy in stellar mass within the halo. By using this definition, the central galaxies make the $\sim$80 per cent of the total sample. 

The Y07 group catalogue includes by construction the halo (virial) mass, $M_h$, 
even when there is only one galaxy in the halo.  
The group masses estimated with this method can recover on average the true halo masses,
with no significant systematics \citep{Yang+2008}. 
The available halo masses in this catalogue are based on either the ranking of the characteristic stellar mass or the ranking of the characteristic luminosity in the group.
We use the halo mass based on the characteristic stellar mass, which is defined
as the sum of
the stellar mass of all the galaxies in the halo with 
$M_r - 5$ log$(h) \leq -19.5$, 
where $M_r$ is the absolute magnitude in the $r$-band.
%}
%However, 
When the central galaxy (usually the most massive galaxy in the halo)
is fainter than the magnitude limit of the sample, the host halo does not have a mass estimate. For this reason, the lower halo mass limit in the Y07 catalogue
is $\sim 10^{11.6} h^{-1}$ $M_{\odot}$.  Furthermore, groups with strong survey edge effects do not have halo mass estimates
regardless the luminosity of their members. The latter affects only 1.6 per cent of all Y07 groups.

We select the central galaxy as the 
one with the highest stellar mass in the halo.
%most massive galaxy in stellar mass within the halo. 
We use a volume-limited sample of central galaxies
out to $z = 0.1$. 
At this redshift, the sample is complete down to the $r$-band absolute magnitude $M_r -$ 5 log$(h) \le -19.6$. 
In addition, we identify those galaxies with estimates of the host halo mass in the Y07 catalogue and
with
measured luminosity-weighted ages of 
%Gallazzi et al. (2005) 
\cite{Gallazzi05}
since this latter parameter is essential to this work.\footnote{Luminosity-weighted ages are obtained from\\ \mbox{http://www.mpa-garching.mpg.de/SDSS/DR4/Data/stellarmet.html}\\ which includes the DR4
version of the tables used in \cite{Gallazzi05}.}
There are 66,692 central galaxies with estimates
of stellar age and host halo mass that satisfy our requirements of absolute magnitude and redshift. %(Table \ref{tabla_DR4_centres}).

The aim of this work is %in order 
to detect the assembly bias effect in central galaxies
by measuring cross-correlation functions %using tracers.
%The sample of tracers is obtained from
against all the SDSS galaxies from the Y07 catalogue
using $M_r -$ 5 log$(h) \le -19.6$ and
0.01 $\leq$ $z$ $\leq$ 0.1 
%the restrictions in $M_r$ and redshift %($\sim$100,000  galaxies). 
(100,244  galaxies).
%as shown in Table \ref{tabla_DR4_tracers}.
%Fig. \ref{mag_z} shows %the relationship among 
%the absolute magnitude in the $r$-band, $M_r$, as a function of redshift 
%with the choice of volume-limited samples,
%which avoid 
%observational bias.

%%%subsub-SECTION
\subsubsection{Age parameter $\delta_t$}
\label{section_age_parameter}

\begin{figure}
\leavevmode \epsfysize=8.9cm %\epsfbox{plots/StAge_Mr_SDSS_together.ps}
\epsfbox{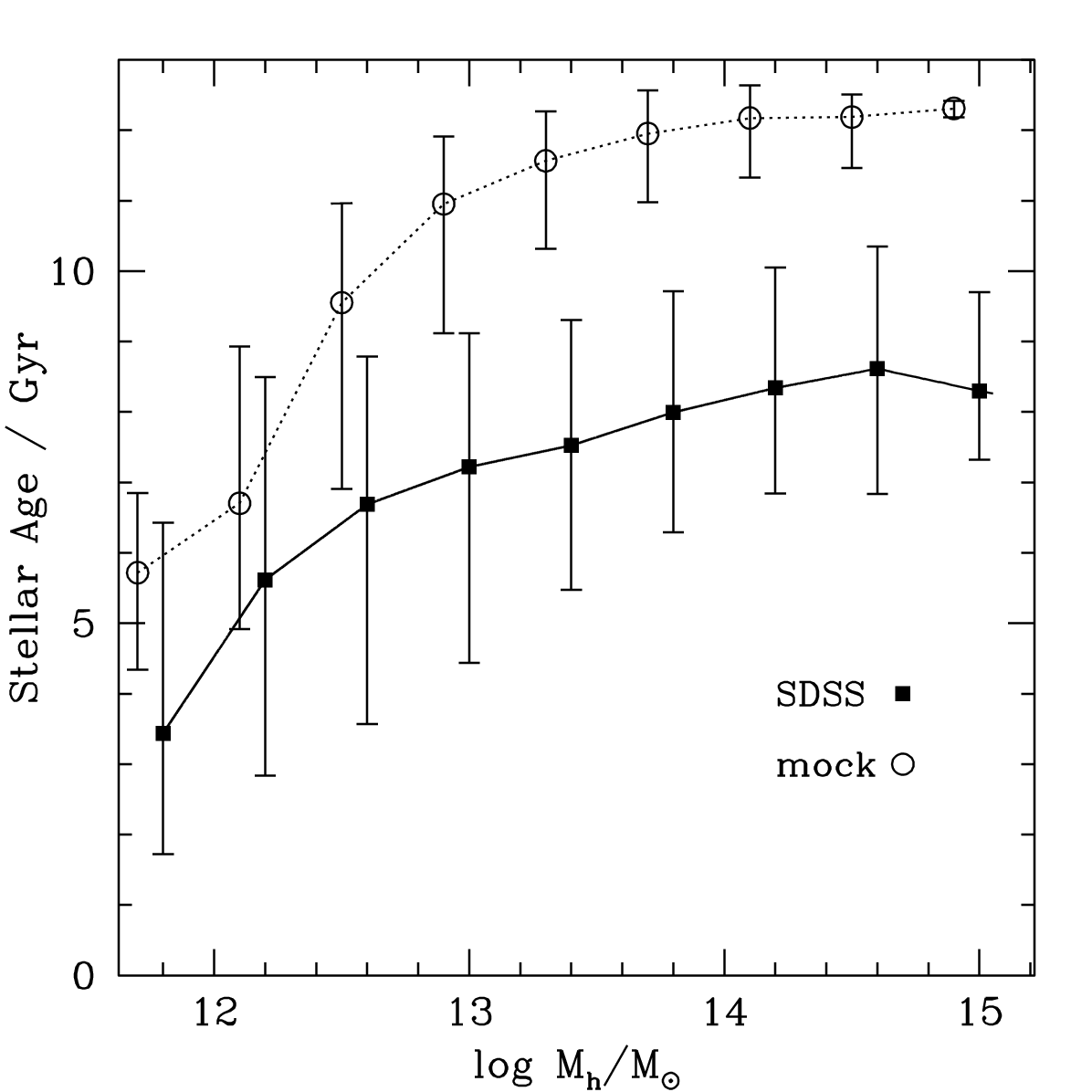}
\caption{
Median luminosity-weighted stellar age 
for the volume-limited sample of SDSS central galaxies (solid squares),
and median mass-weighted stellar age 
for the volume-limited sample of mock central galaxies (open circles),
as a function of the host halo mass ($h = 0.7$).
% and  respectively. 
Error bars correspond to the
10 and 90 percentiles of the stellar age distribution.
For both samples, the median population of galaxies in low-mass haloes is younger
than that of galaxies in massive haloes. 
}
\label{StAge_Mr}
\end{figure}

As was the case %well as 
in Paper I, 
one of the most important parameters throughout this work is the age.
For SDSS galaxies, we use the luminosity-weighted age (or, simply, stellar age).
% \cite{Gallazzi05}. 
%and the $D_n4000$ index (Balogh et al. 1999).
%The latter is defined as
%\begin{equation}
%D_n4000 =  \frac{\overline{F_\nu}(4000 - 4100 \textrm{ \AA})}{\overline{F_\nu}(3850 - 3950 \textrm{ \AA})} ,
%\end{equation}
%\\
%where $\overline{F_{\nu}}$ is the average flux density in the %specified bands (in the rest frame),
%and it is smaller for young stellar populations than %that 
%for old, metal-rich galaxies.
%In our samples there are 76,295 galaxies with estimates
%of stellar age (Table \ref{tabla_DR4_centres}) and 
%205,076(???) with $D_n4000$, both in the magnitude range
%$-21 < M_r < -18$.
To study the assembly bias in galaxies, 
where old objects may have a %higher 
different clustering than young objects of the same mass,
it is not convenient to work directly with the
stellar age, for example, because %they 
this quantity correlates with the mass. %or the luminosity of galaxies.
%We show that, 
For instance, older central galaxies are located, on average,
in massive host haloes (see Fig. \ref{StAge_Mr}).
%as 
%indicated by their luminosity-weighted age.

%%%Fig1

In Paper I, we presented a definition of age which is independent of the 
host halo mass (\mh) %or the luminosity 
through  
the $\delta_t$ dimensionless parameter
%\begin{equation}
\begin{eqnarray}
\delta_{t_{i}} = \frac{t_{i}- \big<t(M_h)\big>}{\sigma_t(M_h)} ,
\label{eq_delta}
%\end{equation}
\end{eqnarray}
%\\
where for the $i$ th galaxy, $t_{i}$ is its stellar age, 
$\big<t(M_h)\big>$ is the
median stellar age as a function of $M_h$, 
and $\sigma_t(M_h)$ is the dispersion around the median
obtained from the 10 and 90 percentiles of the distribution
(error bars in Fig. \ref{StAge_Mr}).  
Then, central galaxies with positive (negative) values
of $\delta_t$ correspond to old (young) populations with respect to
the median distribution of stellar ages at a given 
host halo mass. 
In particular, throughout the paper we use $\delta_t > 0.5$ to select old galaxies and $\delta_t < -0.5$ to select young galaxies. 
The scatter in the observational stellar age can be higher than
the stellar age obtained in simulated galaxies such as the model described in Section \ref{section_sam}.
However, the same cuts in $\delta_t$ to select old and young galaxies are used in the SDSS and simulated samples.

%%%sub-SECTION
\subsection{The mock catalogue}
\label{section_sam}

We construct the mock catalogue using synthetic galaxies from the model by 
\citet[][see also \citealt{Cora06}]{LCP08},
%Lagos, Cora $\&$ Padilla (2008,
%see also Lagos, Padilla \& Cora 2009),
who combine a cosmological
$N$-body simulation of the concordance $\Lambda$CDM universe
and a semi-analytic model of galaxy formation
to follow the evolution of the barionic component of haloes
accross cosmic history.
The numerical simulation consists in a periodic box of 150 $h^{-1}$ Mpc
on a side
that contains 640$^3$ dark matter particles with a mass resolution of
$\sim$10$^9$ $h^{-1}$ M$_{\odot}$. 
The cosmology used is $\Omega_{tot}$ = 1, $\Omega_m$ = 0.28 (with a baryon fraction of 0.16), $\Omega_\Lambda$ = 0.72, $\sigma_8$ = 0.81, \mbox{$h$ = 0.7,} and spectral index $n_s$ = 0.96.
Dark matter haloes are identified as %self-bound 
structures that contain at least 10 particles using a
friends-of-friends (FOF) algorithm 
\citep{Davis85}.
%(Davis et al. 1985).
Another algorithm 
\citep[SUBFIND;][]{Springel01}
%(SUBFIND; Springel et al. 2001) 
is applied
to these groups in order to find substructures
with at least 10 particles.
Typically, the most massive subhalo is considered the main
subhalo, while the other subhaloes are satellites
within a FOF halo.
The galaxy population in the
semi-analytic model is generated using the merger histories of
dark matter haloes and their embedded subhaloes.
The virial mass for galaxies corresponds to the virial mass of 
their respective DM
substructures.

%%% Fig2
\begin{figure*}
\leavevmode \epsfysize=8cm \epsfbox{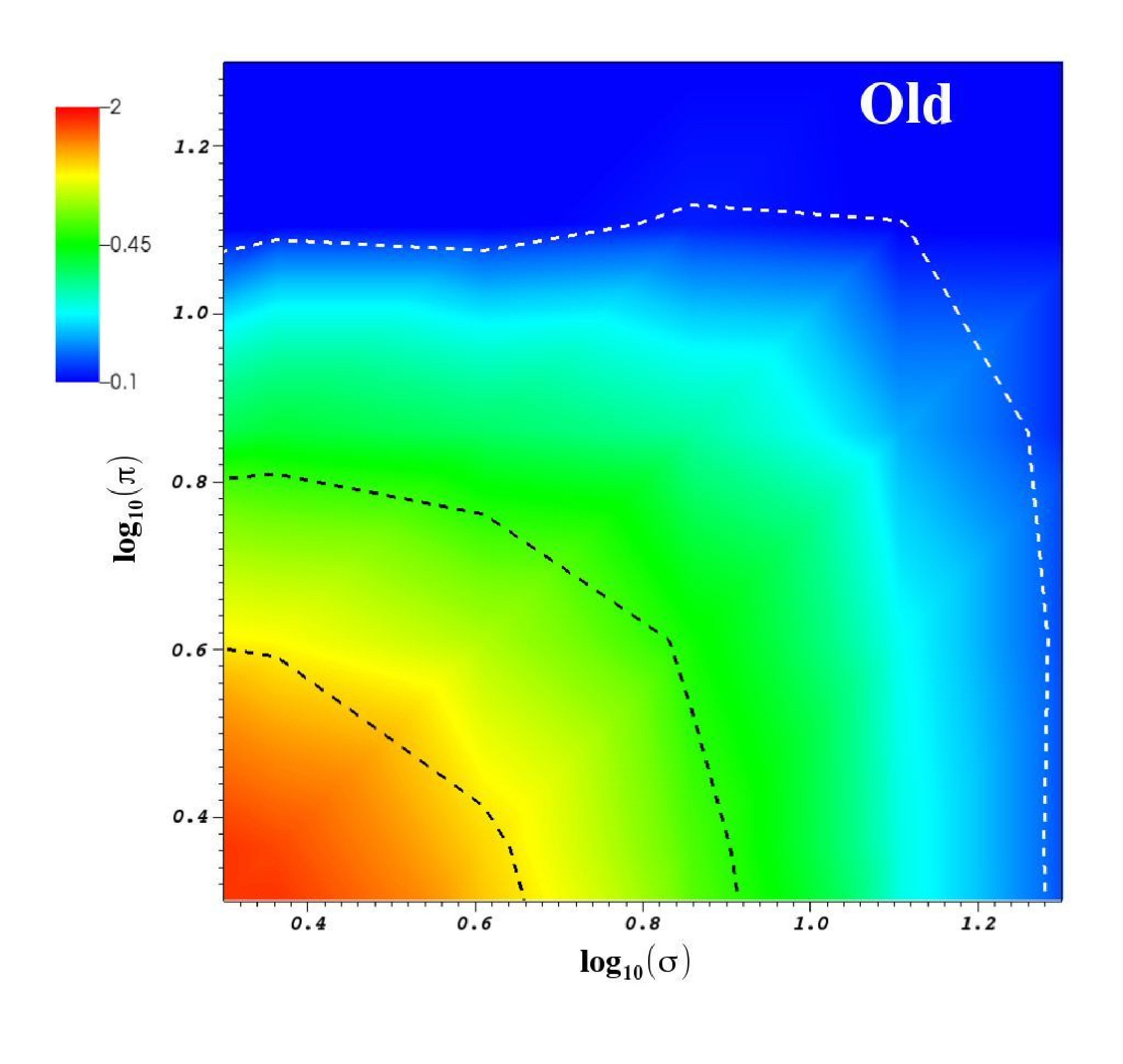}
\leavevmode \epsfysize=8cm \epsfbox{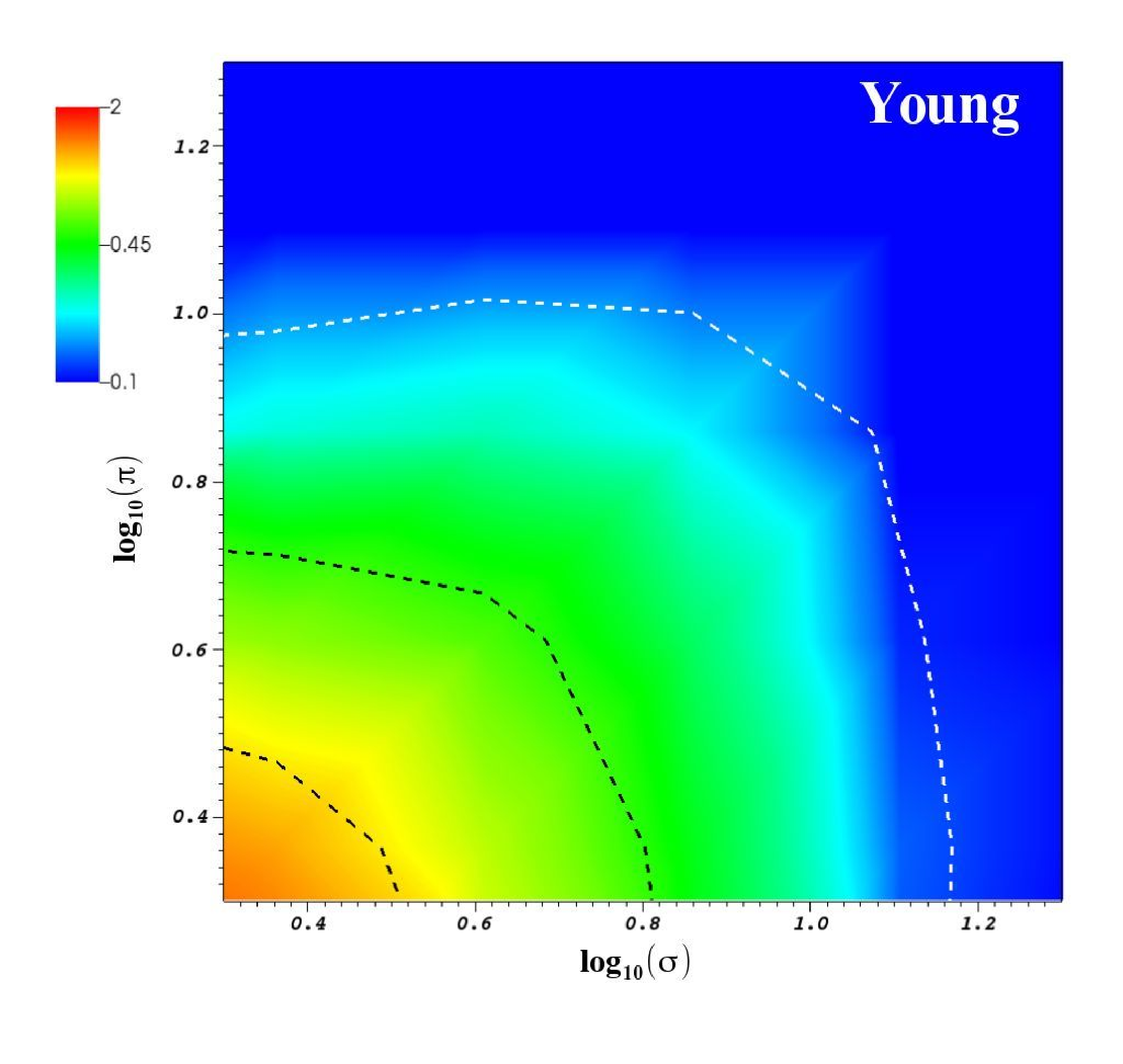}
\caption{
Cross-correlation functions in the directions parallel and
perpendicular to the line of sight for old ($\delta_t > 0.5$, left panel) and young ($\delta_t < -0.5$, right panel) SDSS central galaxies in the host halo mass range  $10^{11.6} < M_h/h^{-1}$ M$_{\odot}$ $< 10^{12}$.
The colour in the figure key indicates the amplitude of the correlation
function. 
The three contour levels, from %top to bottom
large to small separations, correspond to $\xi(\sigma,\pi)$ = 0.1, 0.5 and 1.0, respectively. 
}
\label{Fede_SDSS}
\end{figure*}

In order to make a fair comparison between the simulation and observations, we construct a mock SDSS catalogue using the semi-analytic galaxies.  To do this we place an observer at the position $x,y,z=0$ in the simulation box, and record the positions, peculiar velocities, apparent and absolute magnitudes (at a fixed $z=0$ rest-frame), mass-weighted stellar ages among other properties, for galaxies whose angular positions lie within the observing mask of the SDSS DR4.  The apparent magnitudes are calculated using the comoving distance to each galaxy; we apply an upper limit in magnitudes at $r=17.77$ to mimic the selection function of the %Main Galaxy Sample. 
galaxy sample.
We calculate redshifts taking into account both the comoving distance and the peculiar velocity projected along the line of sight, which implies our mock catalogue shows the Finger-of-God and Kaiser effects.

For the mock central galaxies, we use the virial mass of their host dark matter haloes. In general there is a good qualitative agreement %by using this definition of halo mass and that estimated by 
between the stellar age as a function of halo mass in the mock and %Y07 for 
SDSS central galaxies (see Fig. \ref{StAge_Mr}).
An important difference with respect to the SDSS sample is in the definition of stellar age. In the mock catalogue we use the mass-weighted stellar age in contrast to the luminosity-weighted stellar age of \cite{Gallazzi05}. For this reason, the stellar ages in the mock catalogue appear to be larger than those of the SDSS sample. The relative (normalized) age parameter $\delta_t$ is estimated for the mock central galaxies using  the mass-weighted stellar age in equation (\ref{eq_delta}).
Using the relative age parameter ensures that using the two
different age definitions does not affect our results, since these two age definitions are related via a largely one-to-one relation, even at the low age range where young stars act to further decrease the age of a galaxy.

%%%%%%%%%%%%%%%%%%%%%%%%%%%%%%%%%%%%%%%%%%%%%%%

%%%SECTION
\section{Assembly bias in SDSS galaxies}

%\subsection{Cross-correlation functions for SDSS galaxies}
\label{section_two-point_SDSS}

The trend that old objects are located in regions of higher densities than
young objects of equal mass is %also %present using 
readily seen in
the two-point correlation function, where the ones with the higher clustering
at large scales are those located on higher density regions, 
which could be indicating the locations of
higher initial density peaks.
The correlation function %in real space, $\xi(r)$, 
has the advantage that
it obtains an accurate characterization of the spatial distribution of objects from kpc scales out to
Mpc scales
by simply measuring
the excess of pairs at a given distance %$r$ 
with respect to a random distribution.
Therefore, %$\xi(r)$
the correlation function is a useful tool to study the
large-scale environment.
%The
Our overall results will support that older populations
are located in higher densities compared to younger ones at a given host halo mass.

\begin{figure}
%\leavevmode \epsfysize=8.4cm \epsfbox{plots/DR4_sample1.ps} \leavevmode \epsfysize=8.4cm \epsfbox{plots/DR4_sample2.ps}
%\leavevmode \epsfysize=8.4cm \epsfbox{plots/DR4_sample3.ps} \leavevmode \epsfysize=8.4cm \epsfbox{plots/DR4_sample4.ps}
\leavevmode \epsfysize=8.8cm \epsfbox{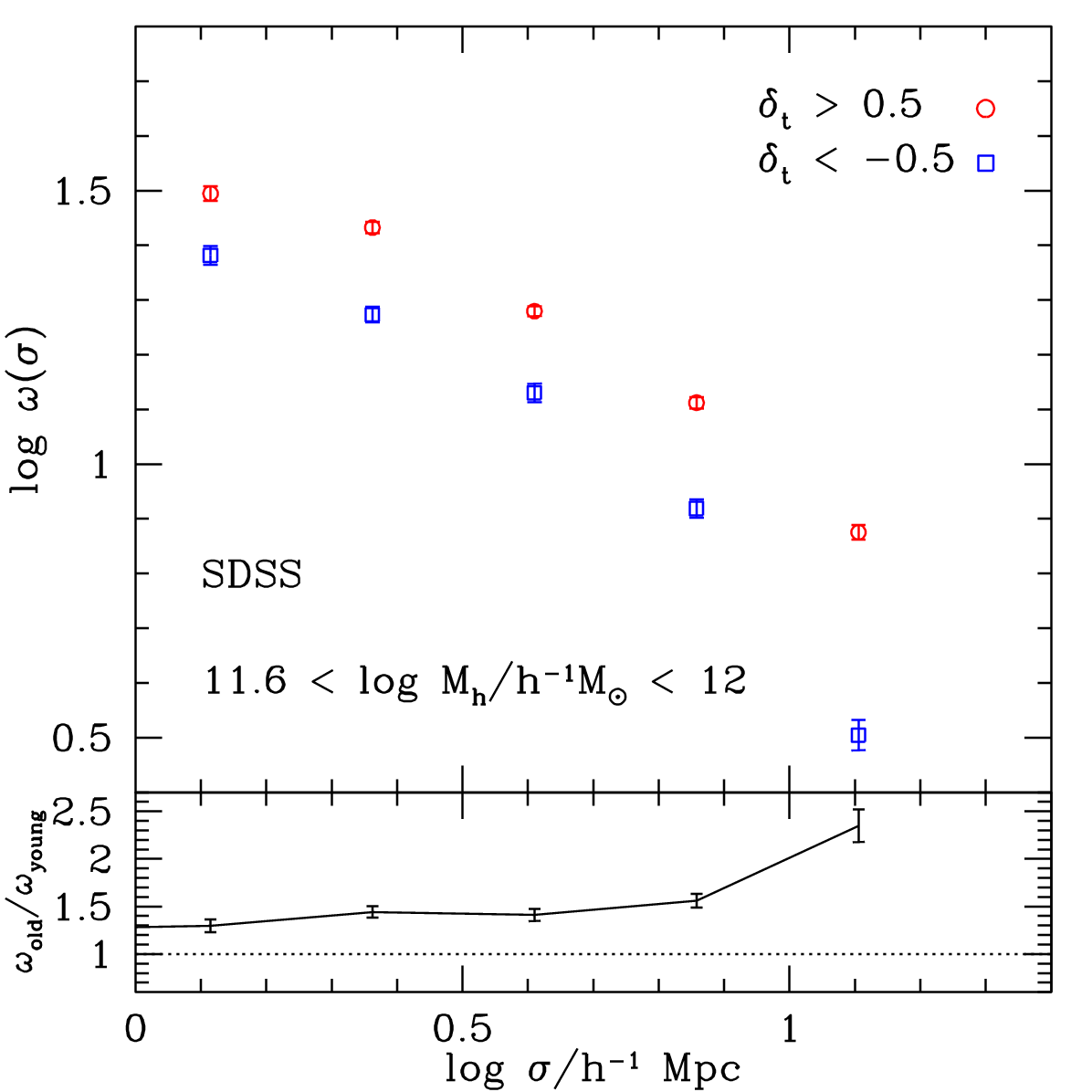} %\leavevmode \epsfysize=8.5cm \epsfbox{plots/fig_xiCL1.ps}
\caption{
Projected cross-correlation functions %(left panel) and cross-correlation functions in real space (right panel) 
for SDSS central galaxies.
The bin in host halo mass $M_h$ is
indicated in %each
the panel. 
Old and young galaxies of the same host halo mass are shown in red circles and blue squares, respectively. 
Error bars are calculated using the jackknife method.
The lower box %in each panel 
corresponds to the
ratio between the correlation functions of the
old and young populations (solid line), where the jackknife errors are taken into account.
The dotted line is the unit ratio (i.e. no assembly bias).
}
\label{xi_DR4}
\end{figure}

In this section, we will test whether %the
SDSS central galaxies show an
assembly-type bias %effect 
by using the 
two-point correlation function. 
%using
%for the samples of centres (central galaxies) and tracers (central$+$satellites galaxies). %for each of 
%in the samples %in 
%of Tables \ref{tabla_DR4_centres} and \ref{tabla_DR4_tracers}.
Due to the redshift space distortions,  
we have to calculate first the cross-correlation in two coordinates
$\xi(\sigma,\pi)$, where $\sigma$ is perpendicular and $\pi$ is parallel
to the line of sight.
Fig. \ref{Fede_SDSS} shows the resulting
$\xi(\sigma,\pi)$ for the old and young central galaxies
in the host halo mass range $10^{11.6} < M_h/h^{-1}$ M$_{\odot}$ $< 10^{12}$.
%$-21 < M_r < -20$. 
As can be seen there is a 
subtle difference between both populations due to 
their slightly offset clustering amplitudes (c.f. Fig. \ref{xi_DR4}),
but also possibly due to different 
finger-of-god and infall effects. %Therefore,
Given that the differences in infall are too subtle, we will not use  these profiles 
to detect the assembly bias in contrast to Paper I.

%%% Fig3

We use the jackknife technique to estimate the variance error in the calculation of $\xi(\sigma,\pi)$, 
which has been shown as a robust error estimator for cosmological samples
\citep{Zehavi02,Cabre07}, although it can overestimate the variance on small scales \citep[][]{Norberg09}.
%2D Cross-correlation function, 
In general, this method consists in dividing the sample in parts and then 
the correlation function is
calculated by discarding one of the subsamples at a time, $\xi_i(\sigma,\pi)$, and
it is compared with the correlation function obtained for the whole sample $\xi(\sigma,\pi)$ as follows
\begin{eqnarray}
\Delta^2 = \frac{1}{N} \sum_{i=1}^N [\xi_i(\sigma,\pi) - \xi(\sigma,\pi)]^2 \textrm {  ,} 
\label{eq_jackknife}
\end{eqnarray}
%\\
where $N$ is the total number of subsamples. In particular,
samples selected from both the SDSS and mock catalogues
%(described in Section \ref{sec_Sim_full})
are divided in 
%ten 
twenty parts, i.e. $N = 20$.

We integrate over the $\pi$ component
to obtain the projected correlation function $\omega(\sigma)$, 
\begin{eqnarray}
\omega(\sigma) = 2 \int_{\pi_{min}}^{\pi_{max}} \xi(\sigma,\pi) d\pi ,
\label{eq_xi_proj}
\end{eqnarray}
%\\
where $\pi_{min}$ = 0.1 $\mpc$ and $\pi_{max}$ = 30 $\mpc$.
%This can be inverted to obtain the real-space cross-correlation function
%using the relation described by 
%\citet[][see also \citealt{Paz08}]{Saunders92},
%\mbox{Saunders, Rowan-Robinson $\&$ Lawrence} 
%(1992, see also Paz, Stasyszyn $\&$ Padilla 2008),
%\begin{eqnarray}
%\xi(r) = \frac{-1}{\pi} \sum_{j \geq i}^{} \frac{\omega(\sigma_{j+1}) - \omega(\sigma_j)}
%{\sigma_{j+1} - \sigma_j} \textrm{ln}\left(\frac{\sigma_{j+1} + \sqrt{\sigma_{j+1}^2 - \sigma_i^2}}
%{\sigma_j + \sqrt{\sigma_{j}^2 - \sigma_i^2}}\right).
%\label{eq_xi_inverted}
%\end{eqnarray}
%\\
As mentioned above, we calculate projected cross-correlation functions using central galaxies. In this case, the clustering amplitude is smaller than using all the galaxies (centrals and satellites) as the main sample  \citep[see e.g.][]{WangL13}.
Furthermore, the projected cross-correlation functions for central galaxies with halo masses larger than  $10^{11.6} h^{-1}$ $M_{\odot}$
are noisy at small scales (i.e. the one-halo term), so that
we are only interested in the estimation of $\omega(\sigma)$ at large scales (i.e. the two-halo term).
Fig. \ref{xi_DR4} shows $\omega(\sigma)$ %and $\xi(r)$, left and right panels respectively, 
for old and young SDSS central galaxies selected using limits in 
$\delta_t$
of equal host halo mass. 
We systematically find that 
the old galaxies show a higher clustering relative
to the young population at large scales ($> 1$ $h^{-1}$ Mpc).
Therefore, SDSS central galaxies of equal host halo mass show a 
significant dependence of clustering amplitude with the age, i.e. the assembly bias effect
%since the
with average differences of 61 $\pm$ 9 per cent
for  $\omega(\sigma)$.\footnote{The uncertainties are estimated by using the jackknife errors.} %and 31\% $\pm$ 5\% for $\xi(r)$.
%%%
% f = mean_ratio +- mean(delta_ratio) 
%%%
%As can be seen from Table \ref{tabla_mean_Mvir_Mr}, the mass-to-light ratio $M/L$
%is not higher for old galaxies at fixed luminosity in the simulation, so that this
%dependence of clustering amplitude with age is the assembly bias effect.
This result is in %conflict with some
agreement with %Wang et al. (2008), Zapata et al. (2009) and Cooper et al. (2010) 
\cite{WangL13}
who find an assembly-type bias in galaxies but using sSFR in bins of stellar mass instead of using stellar age in bins of host halo mass as done in our case.
Our sample definition resembles directly those used to detect the assembly bias in simulations (see e.g. Section \ref{section_two-point_sim}), and therefore constitutes a direct assembly bias measurement.

%%%Fig4
\begin{figure}
%\leavevmode \epsfysize=8.4cm \epsfbox{plots/xi_sim1.ps} \leavevmode \epsfysize=8.4cm \epsfbox{plots/xi_sim2.ps}
%\leavevmode \epsfysize=8.4cm \epsfbox{plots/xi_sim3.ps} \leavevmode \epsfysize=8.4cm \epsfbox{plots/xi_sim4.ps}
\leavevmode \epsfysize=8.8cm \epsfbox{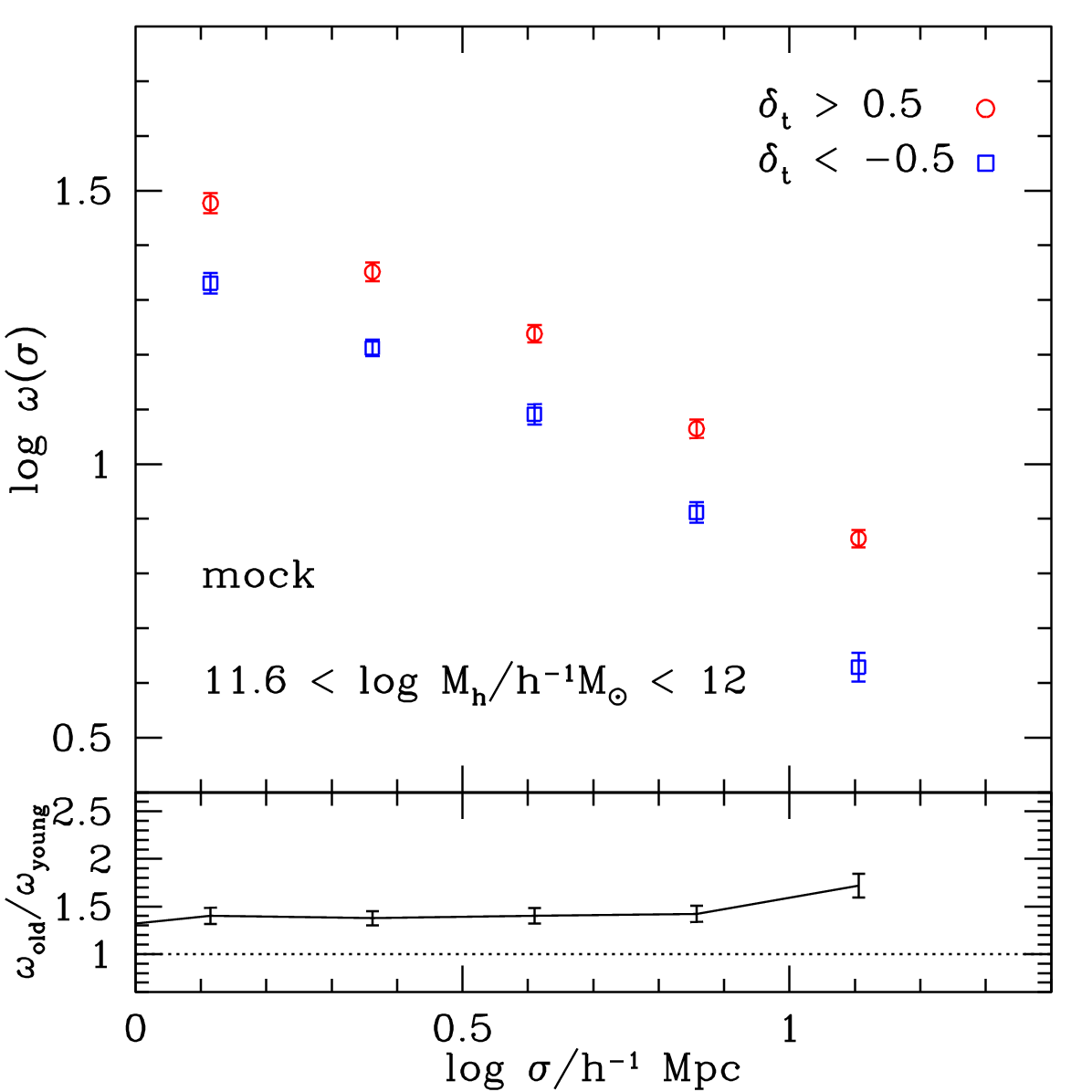} %\leavevmode \epsfysize=8.4cm \epsfbox{plots/figCL1_mock.ps}
\caption{
Same as Fig. \ref{xi_DR4} but for
the mock central galaxies.
}
\label{xi_sim}
\end{figure}

%%%
%\subsubsection{Cross-correlation functions for the simulated galaxies}
\subsection{Assembly bias in the mock catalogue}
\label{section_two-point_sim}

Fig. \ref{xi_sim} shows the 
results of the projected cross-correlation functions 
%in projected and real space 
for the mock central galaxies
hosted by haloes of the same mass as the SDSS samples.
Error bars 
are calculated using the jackknife method. As can be seen, %from these panels, 
old mock central galaxies (red circles) are more strongly
clustered than the young mock central galaxies (blue squares) in haloes of equal mass, with average differences of
46 $\pm$ 9 per cent
in the clustering amplitude of $\omega(\sigma)$
at large scales ($>$ 1 $h^{-1}$ Mpc). 
%and $\xi(r)$, respectively.

The results from observed SDSS central galaxies and mock central galaxies are qualitatively similar;
old galaxies cluster more than young galaxies in host haloes of equal (low) mass.
Quantitatively, 
%in the projected cross-correlation function $\omega(\sigma)$, 
the differences found in the clustering amplitude, 
i.e. the average ratio between $\omega(\sigma)$ of the old and young central galaxies,
of the SDSS and mock samples are %well within the errors
consistent with each other
(e.g. bottom panels of Figs. \ref{xi_DR4} and \ref{xi_sim}). 
We measure a high signal of assembly bias when using $\omega(\sigma)$, 
which is remarkable for the SDSS central galaxies ($50-70$ per cent).

High density environments play a key role in properties such as the formation age. Later in this work we will use this density-age connection
to obtain a peak-height proxy
that traces the assembly bias effect.
We will show that the crowding
around objects helps us trace the assembly bias effect as efficiently as with the smooth density field approach introduced
in Paper I.

%%%%%%%%%%%%%%%%%%%%%%%%%%%%%%%%%%%%%%%%%%%%%%%%%%%%%%%%%%%%%%%%%
\section{Overdensity peak height proxy using relative age and halo mass }
\label{sec_prm}

In Paper I, we presented an approach to trace the assembly bias effect.
This consisted in measuring the total mass inside spheres of 
different radii around semi-analytic galaxies, which in some cases could be larger 
than the virial mass of the host dark matter halo. Using two free parameters
%for the radius, 
to introduce a dependence of this radius on mass and age,
this model defined a new overdensity peak height for which the 
large-scale clustering of objects of a given mass did not depend on the age. 
The peak height can be thought as the size of
the density perturbation that will collapse in a structure, and it is usually well represented by the virial mass; the higher the peak, the more massive will be the structure.
%%%%% EXTRA
We then analyzed the changes of this peak height proxy, compared to virial masses.
We typically found that low-mass, old galaxies have larger peak heights than young galaxies
of equal virial mass, which implies that the environment is more dense for old objects.
They are preferentially surrounded by massive neighbour haloes,
which is in agreement with other works that mention old, low-mass haloes
may suffer a truncated growth due to massive neighbour haloes 
\citep{WangH07,Dalal08,Hahn09}
%(Wang, Mo $\&$ Jing 2007; Dalal et al. 2008; Hahn et al. 2009).
or where additional effects (e.g. tidal forces) can arise in dwarf haloes ejected from massive clusters thus forming earlier than dwarf haloes in the field
\citep{Li13}.
Therefore, the parametrization of the radius of the sphere to measure the environment is a useful tool to understand what is behind the assembly bias.  
%%%%%EXTRA

However, it is difficult to obtain %implement 
the underlying mass information %on 
for real
galaxies. For this reason, the aim of this section is to %do 
design a similar 
procedure, but using the host halo mass of central galaxies from Y07. %Again, we assume 
%We take advantage of the fact that galaxies of equal virial mass %are placed in same magnitude ranges. 
%have similar luminosities (c.f. Section \ref{section_sam}). 
The idea behind our work is not to minimize the bias, but it is to use this method to understand the physical features that produce the assembly bias.

Instead of using spheres as in Paper I, we measure 
the total mass 
%(which correlates with total substructure mass) 
inside a cylinder 
by considering the host halo mass of neighbour galaxies around
%centred on 
each central galaxy 
along the line-of-sight to take into account the limitations imposed by
the redshift space distortions. 
This procedure is similar to the counts-in-cylinders 
\citep[e.g.][and more references therein]{Berrier+11}, but  
the radius
of the cylinder for each central galaxy in $h^{-1}$ Mpc units is parametrized as
\begin{eqnarray}
r = a \textrm{ $\delta_t$} + b \textrm{ log}\left(\frac{M_{h}}{M_{nl}}\right)\textrm{ ,} 
\label{eq_r_prm}
\end{eqnarray}  
%\\
%$L$ is the luminosity of the galaxy and $L_*$ is
%the characteristic luminosity, which both are related with 
%the absolute magnitude
%using $M_r - M_{*} = -2.5$\textrm{ log}$(L/L_*)$, 
where the free parameters are $a$ and $b$, $M_{h}$ is the host halo mass of the central galaxy and $M_{nl}$ is the non-linear mass at $z = 0$.\footnote{It is defined as the mass within a sphere
for which the rms fluctuation amplitude of the linear field is 1.69 times
the critical density of the Universe, which corresponds to the gravitational
collapse in the spherical collapse model.}
As in Paper I, we use log($M_{nl}$/$h^{-1}$ M$_{\odot}$) = 13.38.
($M_{nl}$ is a factor of normalization that only contributes to the size of the radius $r$. The general results after performing the parametrization do not change when modifying the adopted value of the non-linear mass.) 
%characteristic magnitude is $M_{*} = -20.5$ \cite{Zehavi05}.
The length of the cylinder is $\Delta v$ = $\pm$ 500 $\kms$ centred at each central galaxy.\footnote{The choice of this cylinder length corresponds to the radial velocity dispersion of poor galaxy clusters.
Our general results do not change if we use other cylinder lengths, e.g. $\pm$ 750 $\kms$.}
We add the halo mass of neighbour central galaxies (with masses $> 10^{11.6} h^{-1}$ $M_{\odot}$ which is the lower halo mass limit in our SDSS sample) within the cylinder to obtain the total mass of distinct haloes
in this volume. 
%using the relation $M_r - M_{*} = -2.5$\textrm{ log}$(L/L_*)$.
%In case that 
When the radius $r$ is negative, the total mass of the
central galaxy remains unchanged, i.e. $M = M_h$. If this radius is greater than
3 $h^{-1}$ Mpc, %the luminosity is infinite (i.e. 
the galaxy in the center of the cylinder will not be 
considered for further analysis 
(for the best-fitting set of parameters, 
%only the 3.6 percent
none of the central galaxies in the SDSS and mock catalogues
exceed this limit in radius).

Notice that this cylinder will include neighbour central galaxies
and therefore 
this will be %closely 
related to the crowding of the environment of the galaxy, 
a measure of the local density, but with an adaptive volume
size depending on the galaxy age and host halo mass.  
This procedure to
trace the assembly bias will use correlation functions of relatively old and young galaxies to fit the parameters of the cylinder radius,
but this parametrization will only be said to %be 
correlate with the %small-scale 
correction of Paper I 
(missing peak mass) if the excess of
crowding in %old 
galaxies %of equal luminosity, 
correlates with the excess of mass found in %old, low mass 
haloes
reported in our previous paper.
In order to demonstrate that this is the case, we first need to find the parameters that trace the assembly bias using this measurement
of crowding, and then to check whether the relation between crowding and missing peak mass is present.
%At first, we use the semi-analytic galaxies to perform the %redefinition of an overdensity peak height. 
%Later, we will use this model for the SDSS galaxies. 
In contrast to Paper I, we do not use the infall velocity profile because the velocity inferred from redshift space distortions is noisy compared to the velocity from numerical simulations.
We apply the method first to mock galaxies and then to SDSS galaxies.

%%%sub.sec
\subsection{Mock galaxies}
\label{sec_SAMgalx}

When measuring the total mass in each cylinder
in the mock catalogue, 
we use central galaxies with $M_h >  
10^{11.6}$ $h^{-1}$ M$_{\odot}$, which corresponds
to the lower limit in halo mass of the SDSS sample.

We 
select old and young central galaxies
with the 
relative age, $\delta_t$, as measured
in Section \ref{section_age_parameter}
but using the total mass $M$.  
The samples are split into three total mass bins in log($M$/$h^{-1}$ M$_{\odot}$)
with the same width.
For a given mass bin, %each
the old and young subsamples contain the same number of %old and young 
central galaxies. %with $|\delta_t| > 0.5$ 
The cross-correlation functions are obtained in redshift space, $\xi(\sigma,\pi)$,
between the subsamples of old/young mock central galaxies  
%and
against 
the mock sample of %tracers 
centrals and satellites galaxies
that satisfy  
$M_r -$ 5 log$(h) \le -19.6$
and 0.01 $\leq$ $z$ $\leq$ 0.1.
We estimate their projected %and real space counterparts counterpart
correlation functions as described in Section \ref{section_two-point_SDSS}.

In order to find the best-fitting parameters that
do not show an age dependence of the clustering, 
the reduced $\chi^2$ for the correlation 
function statistics 
is defined as
%\begin{eqnarray}
%\chi^2 = \frac{1}{3} \sum_{i}^3 \left(\frac{1}{n_{dof}} \sum_r \frac{ \big[N_{young}(r) - N_{old}(r) \big]^2}{\Delta^2_{N(r)}}\right)_i 
%\textrm{ ,}
%\label{eq_chixi}
%\end{eqnarray}                                                                          
%\\
\begin{eqnarray}
%\lefteqn{
\chi^2 = \frac{1}{3} \sum_{i}^3 \left(\frac{1}{n_{dof}} \sum_r \frac{ \big[\omega_{old}(\sigma) - \omega_{young}(\sigma) \big]^2}{\Delta^2_{\omega}}\right)_i %+} \nonumber\\ 
%& \quad & {} +
%\frac{1}{3} \sum_{i}^3 \left(\frac{1}{n_{dof}} \sum_r \frac{ \big[\xi_{old}(r) - \xi_{young}(r) \big]^2}{\Delta^2_{\xi}}\right)_i 
\textrm{ ,}
\label{eq_chixi}
\end{eqnarray}
where $\omega_{old}(\sigma)$ 
%and $\xi_{old}(r)$ are the corresponding results
is the result for old galaxies and $\omega_{young}(\sigma)$ 
%and $\xi_{young}(r)$ are those
is that for young galaxies. %respectively.
The error is $\Delta^2_{\omega} = \Delta^2_{\omega_{old}} + \Delta^2_{\omega_{young}}$, %and $\Delta^2_{\xi} = \Delta^2_{\xi_{old}} + \Delta^2_{\xi_{young}}$,
where the first term %in each equation 
is the jackknife error for $\omega_{old}(\sigma)$ %and $\xi_{old}(r)$ 
and the second term is that for $\omega_{young}(\sigma)$.
% and $\xi_{young}(r)$, respectively.
This statistics is estimated
within the range  $4< r/h^{-1}$ Mpc $< 15$,
i.e.
in the two-halo term.
%and is calculated within the range $1< r/h^{-1}$ Mpc $< 20$, $N_{young}$ is the number of neighbours for young central galaxies, 
%whereas $N_{old}$ is  
%the same quantity for old central galaxies. 
%The number of neighbours is defined as
%$N(r) =$ $\big<N_t(r)\big> \xi(r)$ $+ \big<N_t(r)\big>$, 
%where  $\big<N_t(r)\big>$ 
%is the mean number of 
%tracers (see Paper I for details).
%The error is 
%$\Delta^2_{N(r)} = \Delta^2_{N_{young}} + \Delta^2_{N_{old}}$, 
%calculated as the relative error of the number of neighbours.
The symbol $n_{dof}$ denotes the number of degrees of freedom.
The $\chi^2$ value is the average over the three total mass bins.
%This
%particular definition of goodness of fit
%avoids selecting parameters favoured by
%large uncertainties that can
%induce spuriously good fits.  
%%%%%%%%% Aparte del xi(r) estimado usando \sigma, también uso xi(r) a partir de \pi en la marginalización %%%%%%%%%

%%%Fig5
\begin{figure}
\leavevmode \epsfysize=8.8cm \epsfbox{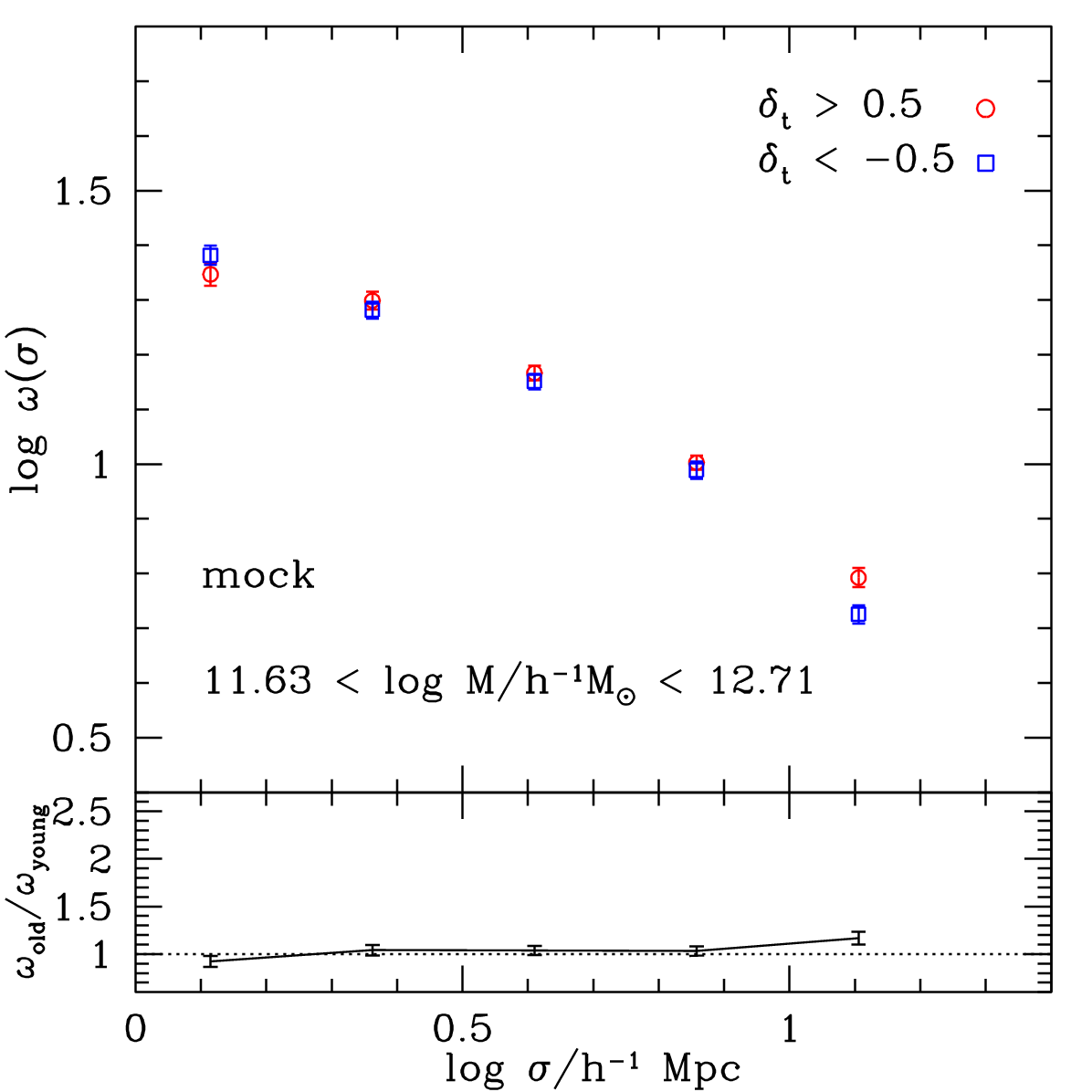}
%\leavevmode \epsfysize=8.4cm \epsfbox{plots/fig_xi1_Sigma_0.3.-0.48.ps}
\caption{
Projected cross-correlation functions 
for mock central galaxies, but with
the new proxy mass $M$ by using the best-fitting values $a$ = 0.3 and $b$ = $-$0.6 in equation (\ref{eq_r_prm}). The mass range is indicated in %each
the panel. Old and young mock galaxies are shown in red circles and blue squares, respectively.
Error bars are calculated using the jackknife method.
}
\label{ratios_sincorr}
\end{figure}

%%%Fig6
\begin{figure}
\leavevmode \epsfysize=8.8cm \epsfbox{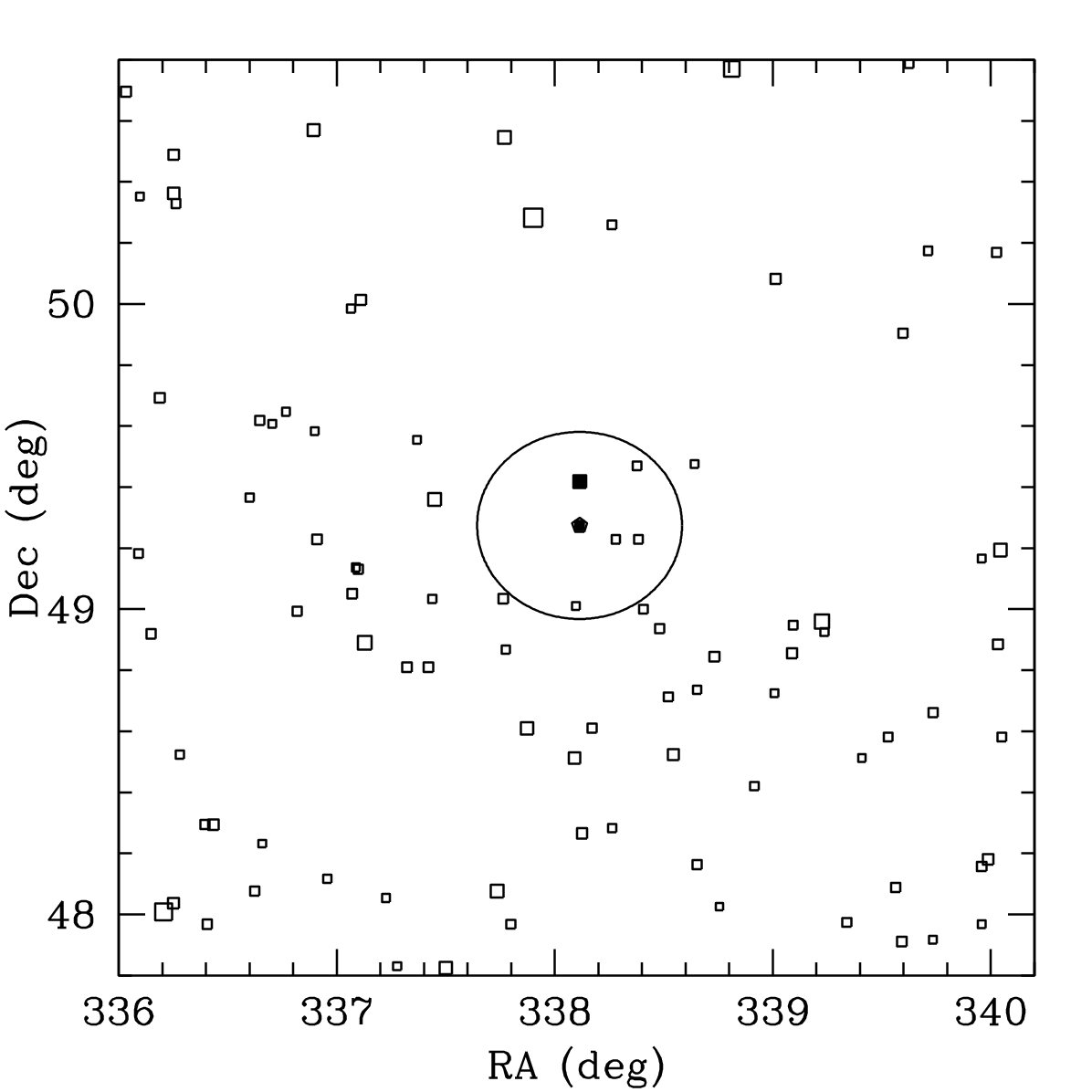}
%\leavevmode \epsfysize=8.4cm \epsfbox{plots/fig_xi1_Sigma_0.3.-0.48.ps}
\caption{Distribution in Right Ascension and Declination of some distinct host halos (squares) in the mock catalogue. The size of the squares is proportional to the halo mass. As an example of the method described in Section \ref{sec_prm}, the filled pentagon shows the position of a central galaxy. %with halo mass $M_h$ = 8.3 $\times$ 10$^{11}$ $h^{-1}$ M$_{\odot}$. 
The solid circle describes the physical radius of the cylinder around this galaxy using equation (\ref{eq_r_prm}) with the best-fitting values $a$ = 0.3 and $b$ = $-$0.6 (see Section \ref{sec_SAMgalx}). The length of the cylinder is $\Delta v$ = $\pm$ 500 $\kms$ with respect to the central galaxy.
The solid squares show all the neighbour distinct host haloes within the cylinder. 
%In this case, it is just one halo of mass $M_h$ = 7.1 $\times$ 10$^{12}$ $h^{-1}$ M$_{\odot}$. The total mass inside the cylinder, or the redefined mass, is $M$ = 7.9 $\times$ 10$^{12}$ $h^{-1}$ M$_{\odot}$.
}
\label{xy}
\end{figure}

%%%Fig7
\begin{figure}
\begin{center}
\leavevmode \epsfysize=4.1cm %6.5cm 
\epsfbox{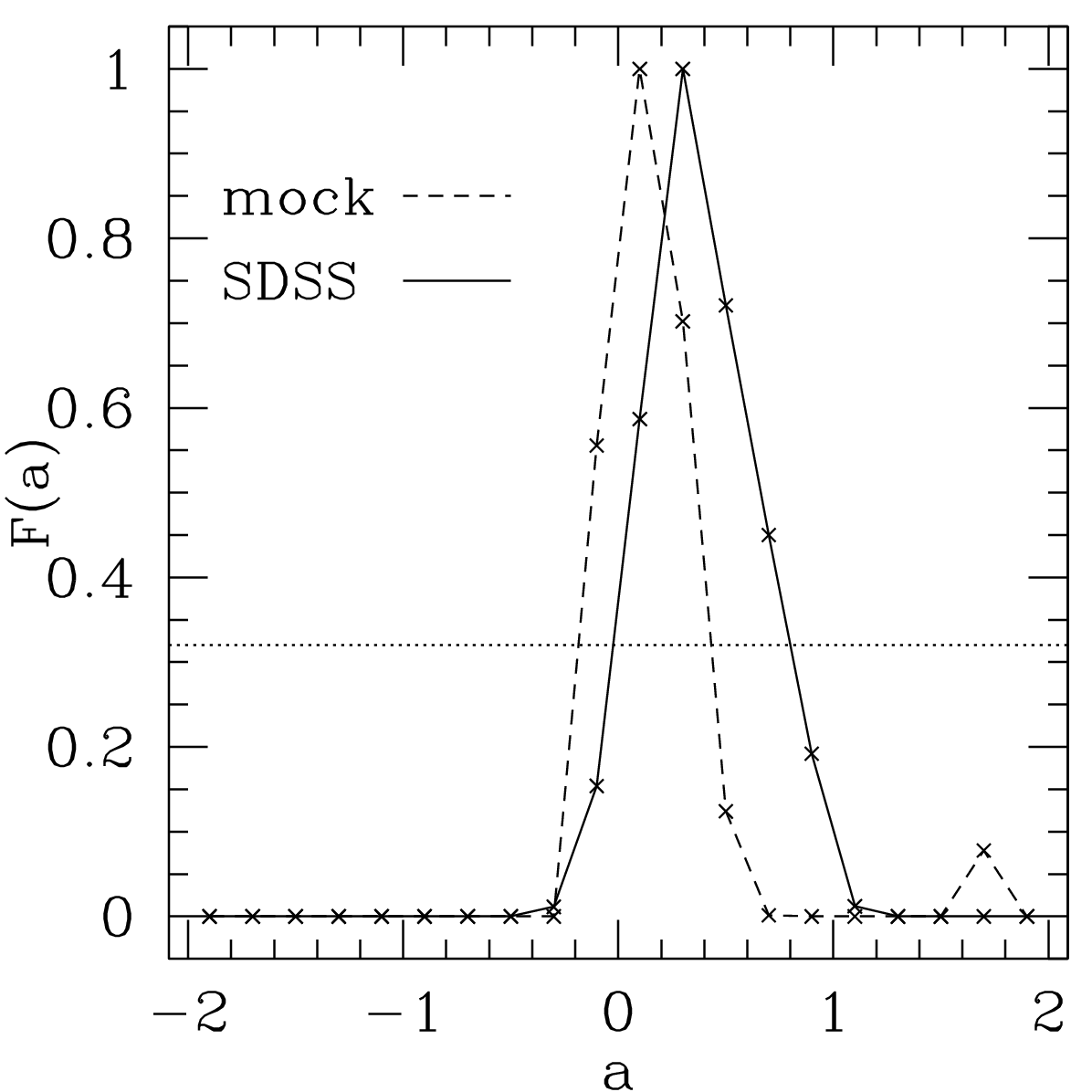} \leavevmode \epsfysize=4.1cm \epsfbox{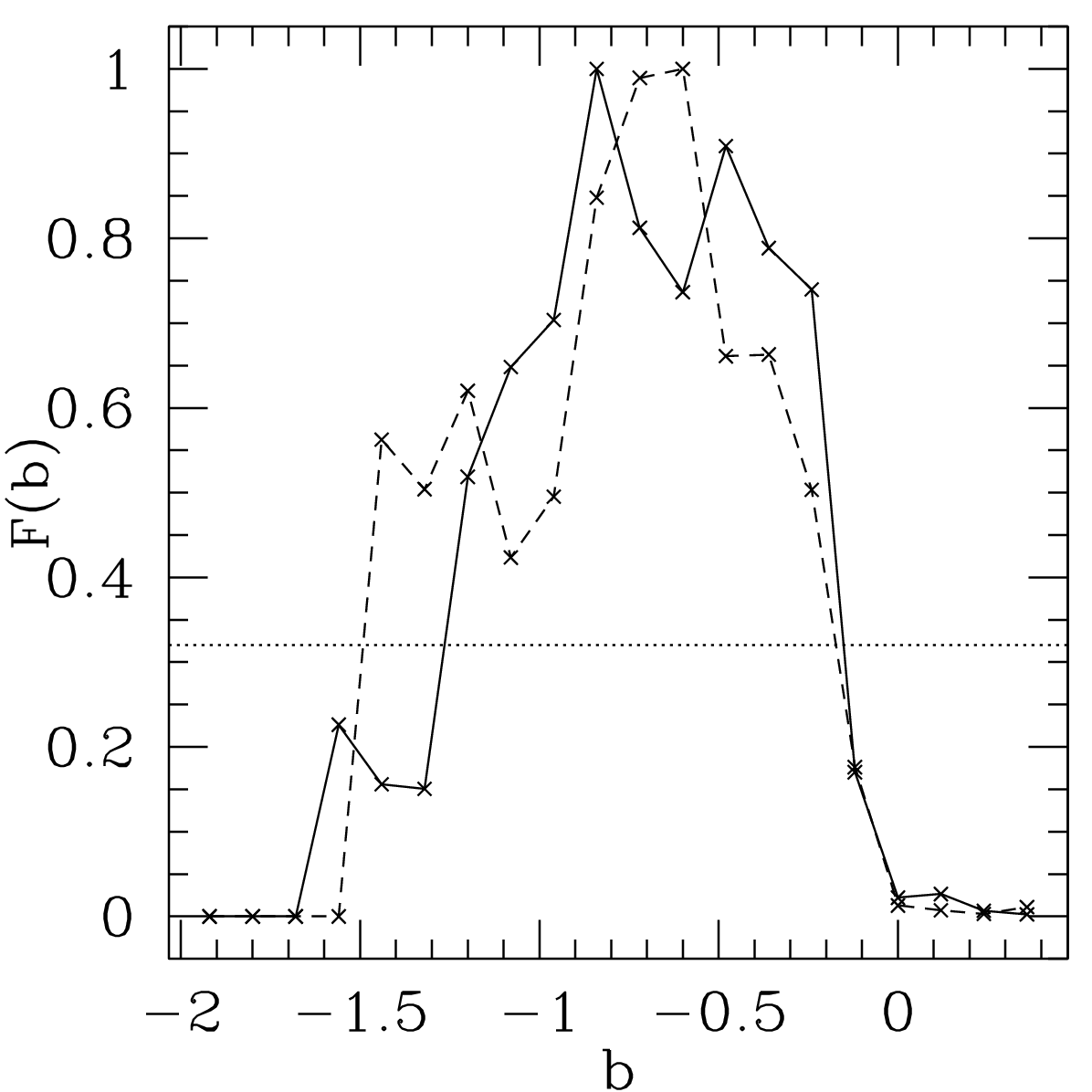}\
\caption{
Marginalized likelihood functions for the free parameters $a$ and $b$ (left and right panels, respectively) using mock galaxies (dashed lines)
and SDSS galaxies (solid lines).
%described in Fig. \ref{comp_Mr_prima} and Section \ref{sec_SAMgalx}, the true total magnitude in
%the $r$-band (dashed lines) and $M'_r$ (solid lines).
Both approaches for the total mass in cylinders of radius $r$
(see equation \ref{eq_r_prm}) obtain roughly similar maximum values in the range
of 1$\sigma$ (intersection of distributions with the horizontal dotted line).
%We also include the results for the magnitude $M'_r$ for SDSS galaxies described in Section \ref{sec_SDSSgalx}
%(dot-dashed lines).
}
\label{likelihoods}
\end{center}
\end{figure}
%%%

The best-fitting parameters corresponding to the highest likelihood in the full parameter space for mock galaxies are
$a$ = 0.3 and $b$ = $-$0.6 ($\chi^2 \sim$ 2.58). 
Fig. \ref{ratios_sincorr} shows %the ratios of 
the cross-correlation functions
of old and young mock central galaxies
using these values, for the lowest mass bin of $M$.
The assembly bias for this mass bin is on average reduced to
4 $\pm$ 6 per cent
%for  $\omega(\sigma)$ 
%and 9\% $\pm$ 5\% for $\xi(r)$
%(left and right panels, respectively) 
%%%practically absent in 
in the mock sample after using this formalism. 
%(compare with solid lines in Fig. \ref{ratios}).
%The average differences in clustering amplitude are smaller than a 10 percent 
%for each range in $M'_r$.

Fig. \ref{xy} shows an example of the method to estimate the total mass. 
The filled pentagon shows the position of a central galaxy with halo mass $M_h$ = 8.3 $\times$ 10$^{11}$ $h^{-1}$ M$_{\odot}$. The solid circle describes the radius of the cylinder around this galaxy using equation (\ref{eq_r_prm}) with the best-fitting values $a$ = 0.3 and $b$ = $-$0.6.
%(see below). 
The solid squares show all the neighbour distinct host haloes within the cylinder. In this case, it is just one halo of mass $M_h$ = 7.1 $\times$ 10$^{12}$ $h^{-1}$ M$_{\odot}$. The total mass inside the cylinder, or the proxy of peak height for the central galaxy (filled pentagon), is then $M$ = 7.9 $\times$ 10$^{12}$ $h^{-1}$ M$_{\odot}$.

%%%Fig8   
\begin{figure*}
\leavevmode \epsfysize=8cm \epsfbox{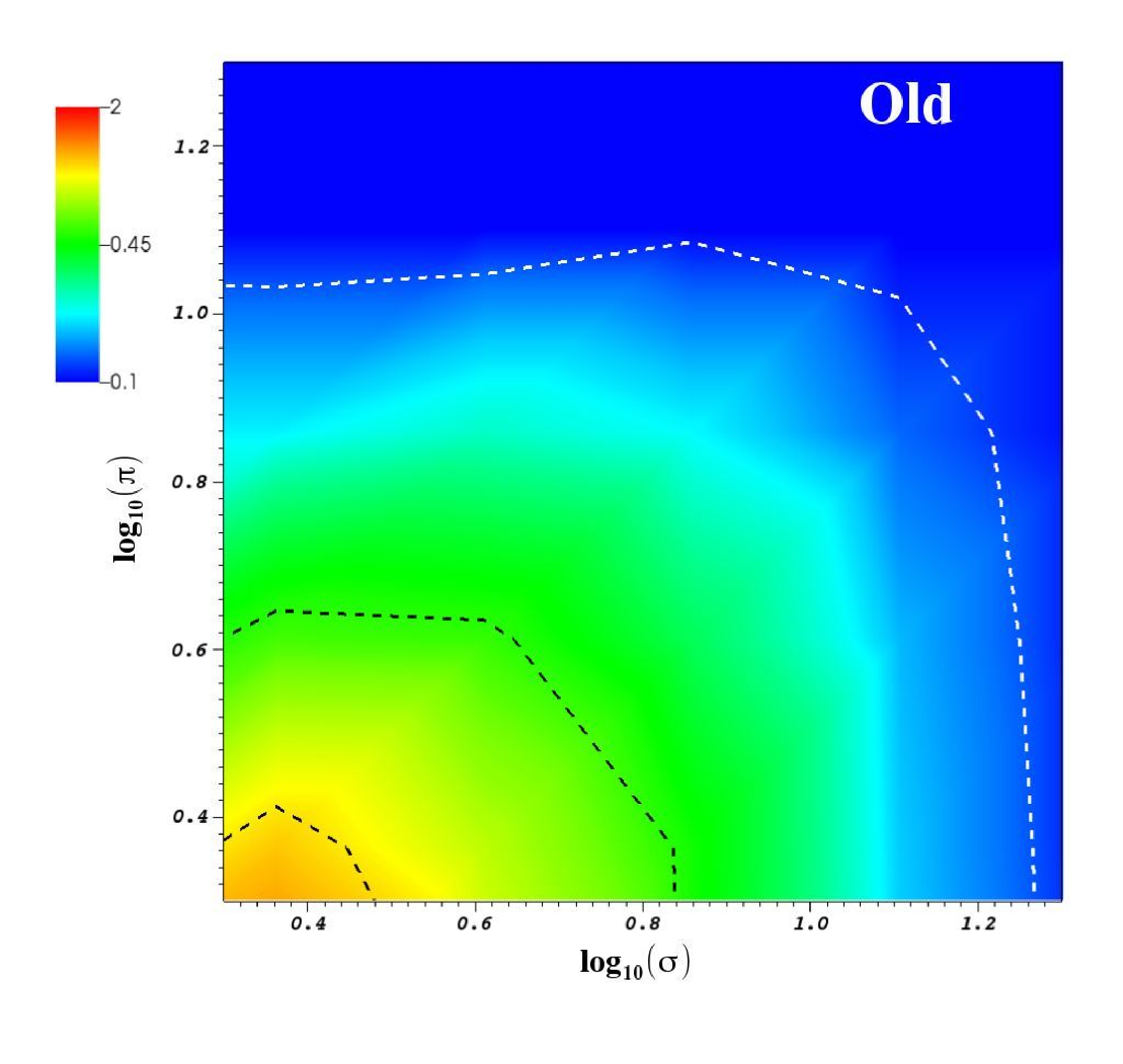}
\leavevmode \epsfysize=8cm \epsfbox{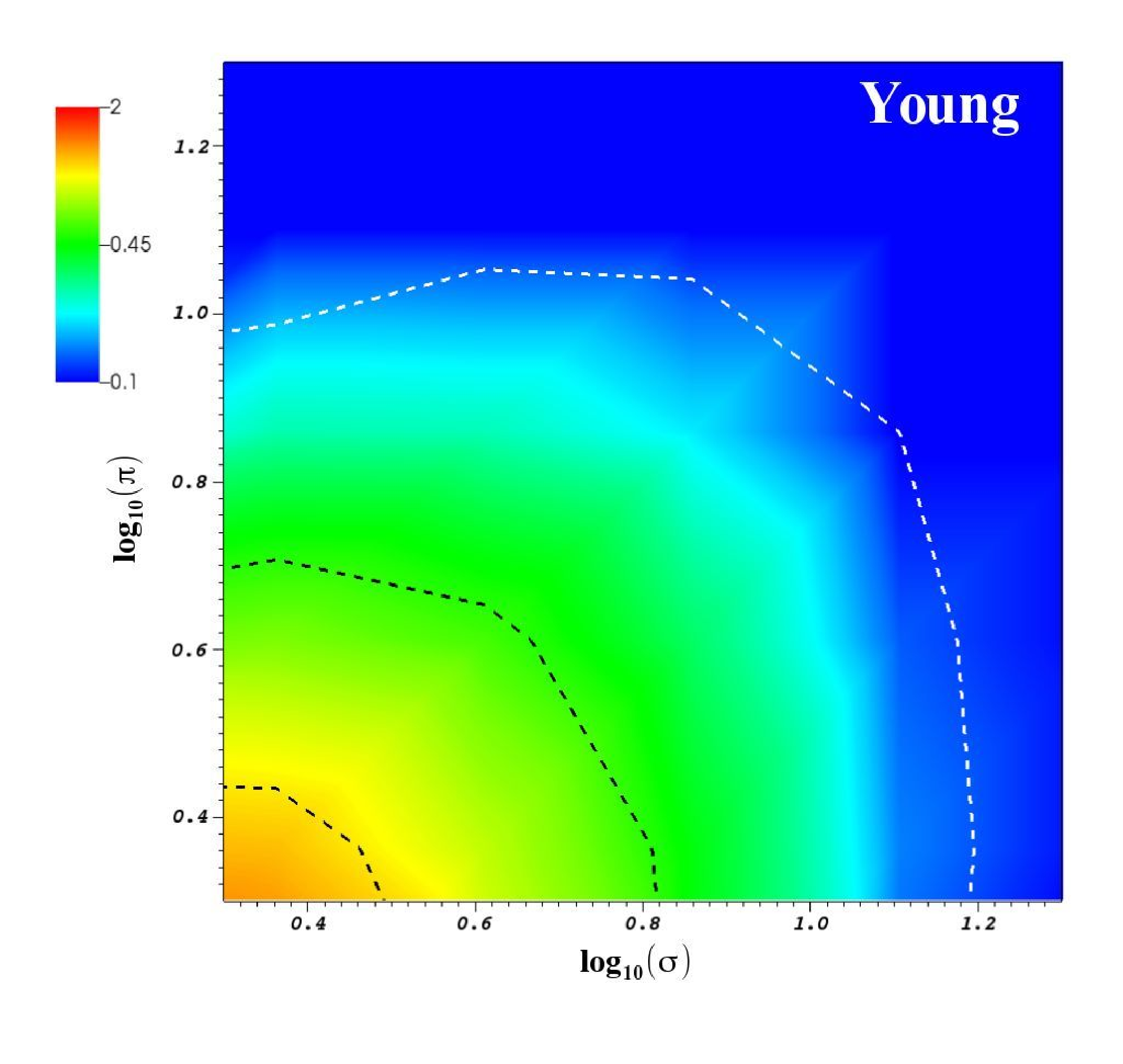}
\caption{
Same as Fig. \ref{Fede_SDSS} also for SDSS central galaxies, but using the proxy for peak height defined by the best fitting parameters
%the best-fitting values 
$a=0.3$ and $b=-1.08$  
in equation (\ref{eq_r_prm}) in the total mass bin %range 
$10^{11.63} < M/h^{-1}$ M$_{\odot}$ $< 10^{12.74}$.
%There are no important differences between old and young populations.
}
\label{Fede_best}
\end{figure*}

Fig. \ref{likelihoods} shows the likelihood functions
for the mock sample (dashed lines) 
and those of the SDSS sample (solid lines, see Section \ref{sec_SDSSgalx} for details). 
Notice that the maximum likelihood in this figure occurs for 
parameter values that do not necessarily coincide with those of
the maximum likelihood in the full parameter space.
We find that 
both approaches for the total mass obtain roughly similar maximum likelihood values
consistent within a 
1$\sigma$ confidence (indicated by the intersections of the
parameter distributions with the dotted line).
%We then 
This justifies our use of $M$ as a reasonable good estimator of the 
%total mass around galaxies % in the range $-22 < M_r < -18$
original peak that gave origin to each central galaxy
with the best-fitting values
$a$ = 0.3 and $b$ = $-$0.6 in the case of the mock catalogue.
%, avoiding the complications introduced
%by the flux limit nature of the observational catalogues.

In the next section we apply the approach to
redefine the overdensity peak height %using luminosities 
%applied to
of SDSS galaxies.

%%%sub-section
\subsection{SDSS galaxies}
\label{sec_SDSSgalx}

Similarly to the previous Section,
we split the samples into three total mass bins in log($M$/$h^{-1}$ M$_{\odot}$)
with the same width.
We select subsamples that contain the same number of old and young SDSS galaxies for a given mass bin, 
with the condition $|\delta_t| > 0.5$, but using the distribution
of the luminosity-weighted stellar age as a function of the total mass $M$.
Recall this mass is estimated inside a cylinder of radius $r$.
%in Equation (\ref{eq_delta}). 
The correlation functions are obtained in redshift space, $\xi(\sigma,\pi)$, %as shown in Fig. \ref{Fede_best}.
and we estimate their projected %and real space counterparts
counterpart as described in Section
\ref{section_two-point_SDSS}.

\begin{figure}
%\leavevmode \epsfysize=8.9cm \epsfbox{plots/1.-0.36_ratios.ps}
\leavevmode \epsfysize=8.8cm \epsfbox{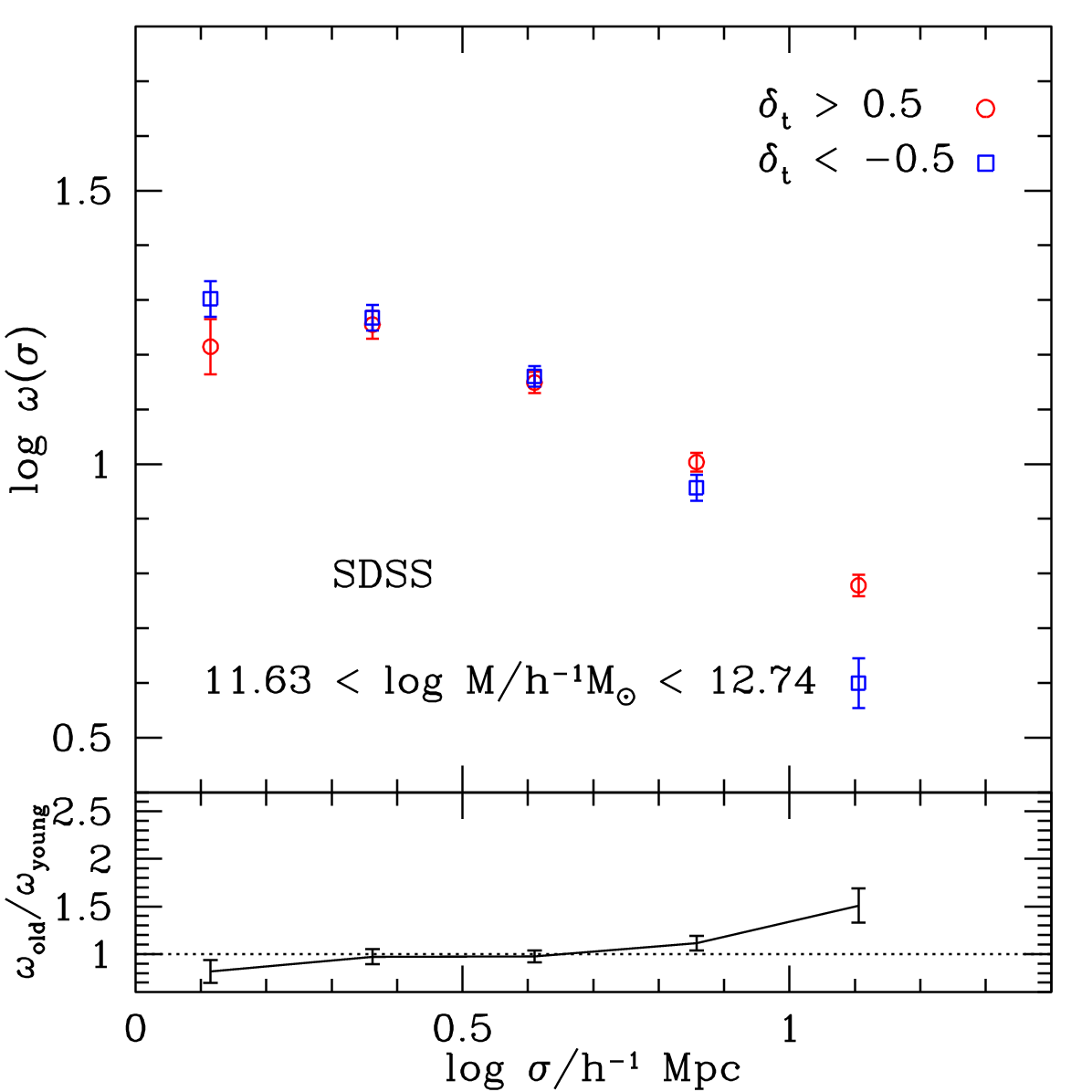}
%\leavevmode \epsfysize=8.4cm \epsfbox{plots/Y_xi1_Sigma_0.3.-0.48.ps}
\caption{Projected cross-correlation functions 
for SDSS central galaxies, but with
the new proxy mass $M$ by using the best-fitting values $a$ = 0.3 and $b$ = $-$1.08. The mass range is indicated in %each
the panel. Old and young SDSS galaxies are shown in red circles and blue squares, respectively.
Error bars are calculated using the jackknife method.
}
\label{sdss_noassembly}
\end{figure}

For the SDSS central galaxies, the best-fitting values
are $a=0.3$, $b=-1.08$ ($\chi^2 \sim 12.35$). 
The marginalized probability distributions for $a$ and $b$
are shown as solid lines in Fig. \ref{likelihoods}, and as can be seen there is a reasonable good agreement with the distributions obtained from our mock catalogue.
%It is remarkable that the
%parameter $a$ has a value only slightly greater compared to that obtained for the galaxies in the simulation
%($a=0.75$), which lies within the uncertainties of this parameter,
%and 
%the value of $b$ is equal in both cases.
The correlation functions in redshift space and the projected correlation functions 
are shown
in Figs. \ref{Fede_best} and  \ref{sdss_noassembly}, respectively, for the lowest mass bin using these values. 
Again, the effect of assembly bias is %almost not present
reduced using the proxy for the overdensity
peak height at scales $ >$ 1 $h^{-1}$ Mpc; for the lowest mass bin the average differences in clustering amplitude are 
8 $\pm$ 10 per cent in the case of $\omega(\sigma)$. 
%and (0.95)\% $\pm$ 5\% for $\xi(r)$ 
%(left and right panels, respectively). 
%smaller than a 10 per cent
%for each range in magnitude $M'_r$),
We confirm that our approach is able to define an initial peak-height proxy using the crowding of the region
where galaxies lie, in a similar manner to that implemented in Paper I, and that SDSS galaxies show a similar behaviour to mock galaxies
since their best-fitting sets of parameters
are comparable.

In this context, as mentioned above, Fig. \ref{likelihoods} shows the marginalized likelihood distributions for the free parameters $a$ and $b$
of equation (\ref{eq_r_prm}) for the SDSS and mock catalogues.
%the new magnitude $M'_r$ for simulated galaxies, the true total magnitude in the $r$-band using also simulated galaxies and $M'_r$ for SDSS galaxies. 
By estimating a weighted average for the values in the range of $1\sigma$ for the two distributions, 
we obtain\\\\
%\begin{displaymath}
%a = 0.76 \pm 0.31 
%\end{displaymath}
%\begin{displaymath}
%b = -0.39 \pm 0.18
%\textrm{ .}
%\end{displaymath}
$ a = 0.26  \pm 0.19$\\ 
$ b = -0.75 \pm 0.36$.\\\\
The best estimates for the $a$ parameter are always positive, whereas
those for the $b$ parameter are negative when defining an overdensity peak height proxy using discrete masses. The next section 
will discuss the properties of this proxy of peaks.
%when using
%host halo masses around central galaxies.

%%%Fig9

%%%%%section
\section{Peaks defined using discrete masses}
\label{sec_peak_lum}

The redefinition of the overdensity
peak height using halo masses within cylinders is able to
trace the assembly bias (Section \ref{sec_prm}).
Table \ref{tabla_radius_mock} shows the median radius of equation (\ref{eq_r_prm})
in 
physical units of $h^{-1}$ kpc 
for three
different bins in $M_h$ using the mock central galaxies. 
The results are shown for the best-fitting values 
$a = 0.3$, $b = -0.6$. The radius of the cylinder for all the mock central galaxies decreases to higher masses.
This is in agreement to 
what was discussed in Paper I,
where the radius decreases to more massive objects.
Bear in mind that the radius can be negative, $r < 0$,
since both the first and second term of equation (\ref{eq_r_prm}) can be negative, which implies  $M = M_h$. 
The results are also split according to
the classification of old and young populations after performing the redefinition of mass for these mock central galaxies.
As in Paper I,
old objects have higher radii than young objects of equal mass $M_h$, thus suggesting stronger
environmental dependences for old populations.
%The median sizes for both sets of parameters are similar for young galaxies, with differences smaller than
%37 $h^{-1}$ kpc, while these differences for old galaxies range from 8 $h^{-1}$ kpc out to 200 $h^{-1}$ kpc 
%toward brighter magnitudes.

%%%mock galaxies
%%%%%%%%%%% TABLA 1
\begin{table*}
%\begin{sidewaystable}
  \centering
\caption[Median radii in physical units
from equation (\ref{eq_r_prm}) for mock central galaxies 
as given by the best-fitting parameters $a = 0.3$, $b = -0.6$.]
{Median radii %in units of virial radius, $r_{vir}$, and 
in physical units ($h^{-1}$ kpc) 
from equation (\ref{eq_r_prm}) for mock central galaxies (see details in Section \ref{sec_SAMgalx}), 
as given by the best-fitting parameters $a = 0.3$, $b = -0.6$.
The results are shown for all mock centrals and, 
also, split among
the old and young populations %of mock central galaxies 
after performing the redefinition of mass.
The radius $r < 0$ implies $M = M_h$.
The halo mass $M_h$ is in units of $h^{-1}$ $M_{\odot}$.
%The ranges in mass $M$ are those shown in Table \ref{tabla_DR4_centres}.
}
\begin{tabular}{c c c c c c}\\

\hline
\hline
Mock: best-fitting parameters            &   Ages  & $\big<r/h^{-1}$kpc$\big>$ & $\big<r/h^{-1}$kpc$\big>$ & $\big<r/h^{-1}$kpc$\big>$ \\

\hline
& & log $M_h$ =  $(12,12.6)$ & log $M_h$ = $(13,13.6)$ & log $M_h$ =  $(14,14.6)$ \\
%\hline
\newline

$a$ = 0.3, $b= -0.6$ & All                  &  700.0        & 111.3     &   $<$ 0  \\
                     & Old                  &  845.9        & 273.9     &   $<$ 0  \\
                     & Young                &  371.0        &  $<$ 0    &   $<$ 0   \\
\hline

\end{tabular}
\label{tabla_radius_mock}
\end{table*}
%\end{sidewaystable}

%%%SDSS Tabla 2
\begin{table*}
%\begin{sidewaystable}
  \centering
\caption[Median radii in physical units
from equation (\ref{eq_r_prm}) for SDSS central galaxies 
as given by the best-fitting parameters $a = 0.3$, $b = -1.08$.]
{Median radii %in units of virial radius, $r_{vir}$, and 
in physical units ($h^{-1}$ kpc) 
from equation (\ref{eq_r_prm}) for SDSS central galaxies (see details in Section \ref{sec_SDSSgalx}), 
as given by the best-fitting parameters $a = 0.3$, $b = -1.08$.
The results are shown for all SDSS centrals and, 
also, split among
the old and young populations %of SDSS central galaxies 
after performing the redefinition of mass.
The radius $r < 0$ implies $M = M_h$. 
The halo mass $M_h$ is in units of $h^{-1}$ $M_{\odot}$.
%The ranges in magnitude are those shown in Tables \ref{tabla_DR4_centres} and \ref{tabla_radius_lum_LCP}.
}
\begin{tabular}{c c c c c c}\\

\hline
\hline
SDSS: best-fitting parameters            &   Ages  & $\big<r/h^{-1}$kpc$\big>$ & $\big<r/h^{-1}$kpc$\big>$ & $\big<r/h^{-1}$kpc$\big>$ \\

\hline
& & log $M_h$ =  $(12,12.6)$ & log $M_h$ = $(13,13.6)$ & log $M_h$ =  $(14,14.6)$ \\
%\hline
\newline

$a$ = 0.3, $b= -1.08$  & All               &  1231.8   & 177.8     &   $<$ 0  \\
                       & Old               &  1465.1   & 376.2     &  $<$ 0   \\
                       & Young             &   971.2   &  $<$ 0    &  $<$ 0   \\

\hline

\end{tabular}
\label{tabla_radius_SDSS}
\end{table*}
%\end{sidewaystable}

Table \ref{tabla_radius_SDSS} shows the median radius of equation (\ref{eq_r_prm})
using the SDSS central galaxies. 
In general, there is a good agreement of the radius sizes with respect to
the mock catalogue (see Table \ref{tabla_radius_mock}) for the massive haloes ($M_h > 10^{13}$ $h^{-1}$ $M_{\odot}$), with differences among them smaller than 102 $h^{-1}$ kpc, %for all the central galaxies,
%102 $h^{-1}$ kpc for the old central galaxies, and 24 $h^{-1}$ kpc for the young galaxies.
but the radius for less massive host haloes ($M_h\sim10^{12.3}$ $h^{-1}$ $M_{\odot}$) in the case of SDSS galaxies
is $\sim600$ $h^{-1}$ kpc higher than the radius in the case of mock galaxies of equal host halo mass.
This means that the best-fitting set of parameters that make the assembly bias %practically absent
very low for both mock and real galaxies
are somewhat similar when using discrete masses in cylinders as a proxy of peak height, although with the need of higher radius for SDSS galaxies in relatively low-mass host haloes.
One reason for this might be related with the fact that, in general, semi-analytic models do not reproduce all the phenomenology of real galaxies (e.g. mass downsizing). However, note that the radius $r$ in equation (\ref{eq_r_prm}) is strongly determined by the free parameters $a$ and $b$. 
%For example, for galaxies with average stellar ages, i.e. $\delta_t \sim 0$, in the low-mass regime,  their radius is $r \sim -b$. According to the right panel of Fig. \ref{likelihoods}, good values ($> 1\sigma$) for the $b$ parameter range from $-0.3$ to $-1.5$. This means radii from $r \sim 300$ kpc to $\sim 1500$ kpc for both SDSS and mock galaxies. 
Therefore, the differences in the radii reported in Tables \ref{tabla_radius_mock} and \ref{tabla_radius_SDSS} are mainly due to the fact that the best-fitting value is $b=-0.6$ for mock galaxies and $b=-1.08$ for SDSS galaxies. %However, as mentioned above, the range $-0.3$ to $-1.5$ in $b$ is acceptable for both samples, which means that an opposite trend could be obtained using other values in this range.
Given the uncertainties (e.g. Fig. \ref{likelihoods}), it is premature to infer any physical interpretation in this regard.

%%%Fig10
\begin{figure}
%\leavevmode \epsfysize=8.9cm \epsfbox{plots/Mr_prima_M9.1.ps}
\leavevmode \epsfysize=8.9cm \epsfbox{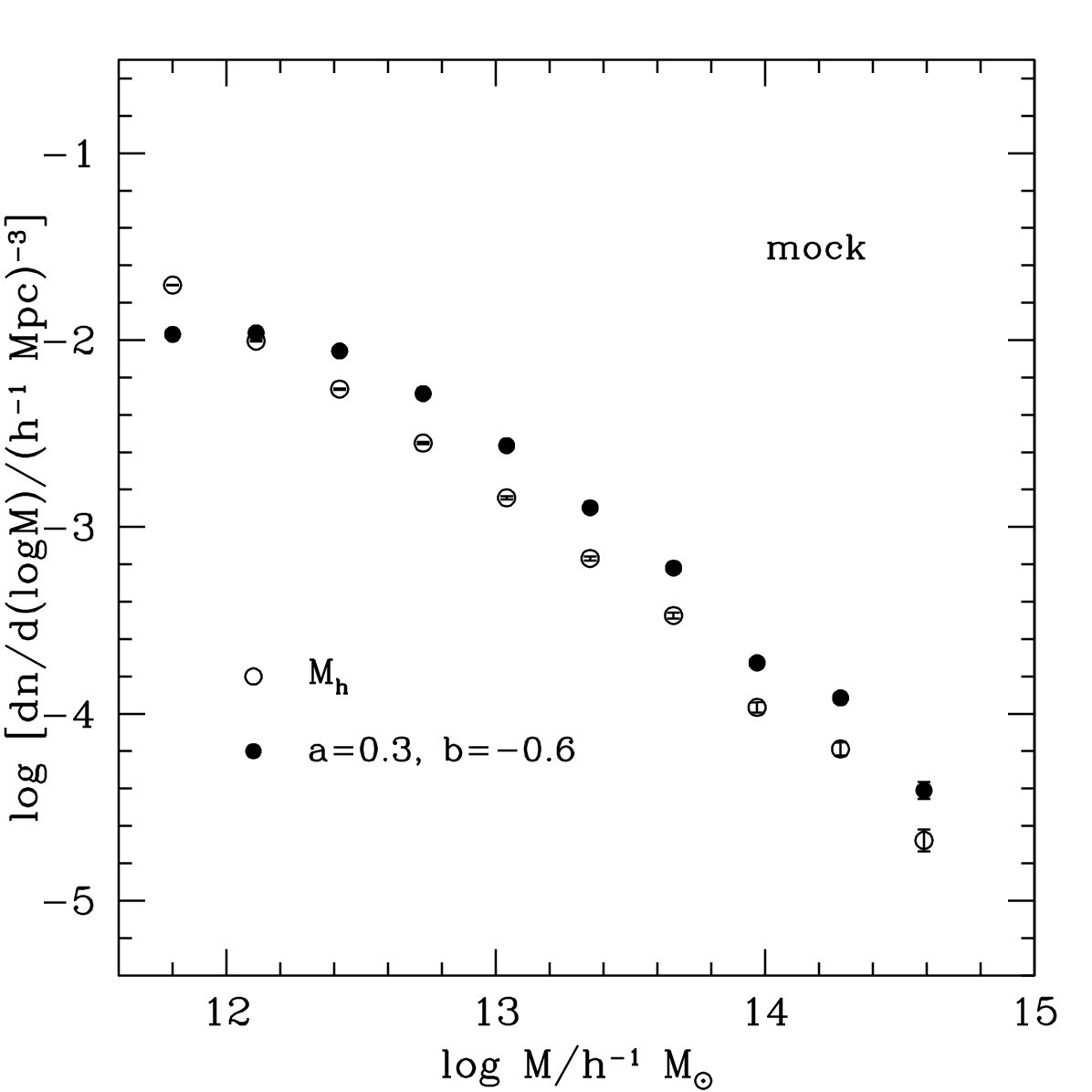}
\caption{
Mass function using the %free
parameters $a=0.3$ and $b=-0.6$ (solid circles) in the mock catalogue. Open circles 
correspond to the mass function using the original host halo mass, $M_h$.
}
\label{MF}
\end{figure}

The method followed to redefine the peak height could be compared
to adding up the 
mass from the smooth density field beyond
the virial radius. 
First, we compare the resulting 
$M$ mass function in the mock catalogue with the original mass function of haloes with mass $M_h$ to help interpret the changes introduced by our approach; this is equivalent to our procedure in Paper I where we compared the mass functions for the original virial mass, and that resulting from our redefinition of peak height.
%although in this
%case the luminosity function does not have a theoretical %prediction.
Fig. \ref{MF} shows %important 
differences in
the mass function
that results %with this model
from the redefinition of mass $M$ for mock central galaxies (solid circles) and that 
from the host halo mass $M_h$ (open circles).
Recall we only use host haloes of central galaxies with the conditions 
$M_r -$ 5 log$(h) \le -19.6$
and 0.01 $\leq z \leq 0.1$.
One
could have expected 
the number of these objects with mass $M$ %is
to be similar
to that of all haloes of mass $M_h$.
%but 
The difference between the
mass functions measured with $M$ and $M_h$
is relatively small but not negligible.  
In the next section we will study %if
the origin of this %issue. 
discrepancy.

%%%sub-section
\subsection{Discrete mass tracers compared to underlying smooth density field}
\label{sec_comp_UnderDis}

In Paper I we found that the redefinition based on the mass density field does not change
the halo mass function at $M > 10^{12}$ $h^{-1}$ M$_{\odot}$ (see their Fig. 8)
when compared to that %of Sheth, Mo, \& Tormen (2001, SMT) or that %using
obtained from the virial mass.
%which
This is in agreement with the result found by 
%Gao, Springel \& White (2005)
\cite{Gao05} where the assembly
bias with respect to formation time is not important for the high mass regime.
The reason for the change found in the low-mass regime 
is that the peak for an old, low-mass object adds more haloes and mass than for a young object of equal mass.

%%%Fig11
\begin{figure}
\begin{center}
%\leavevmode \epsfysize=7cm\epsfbox{plots/comp_masses.ps}
\leavevmode \epsfysize=8.2cm\epsfbox{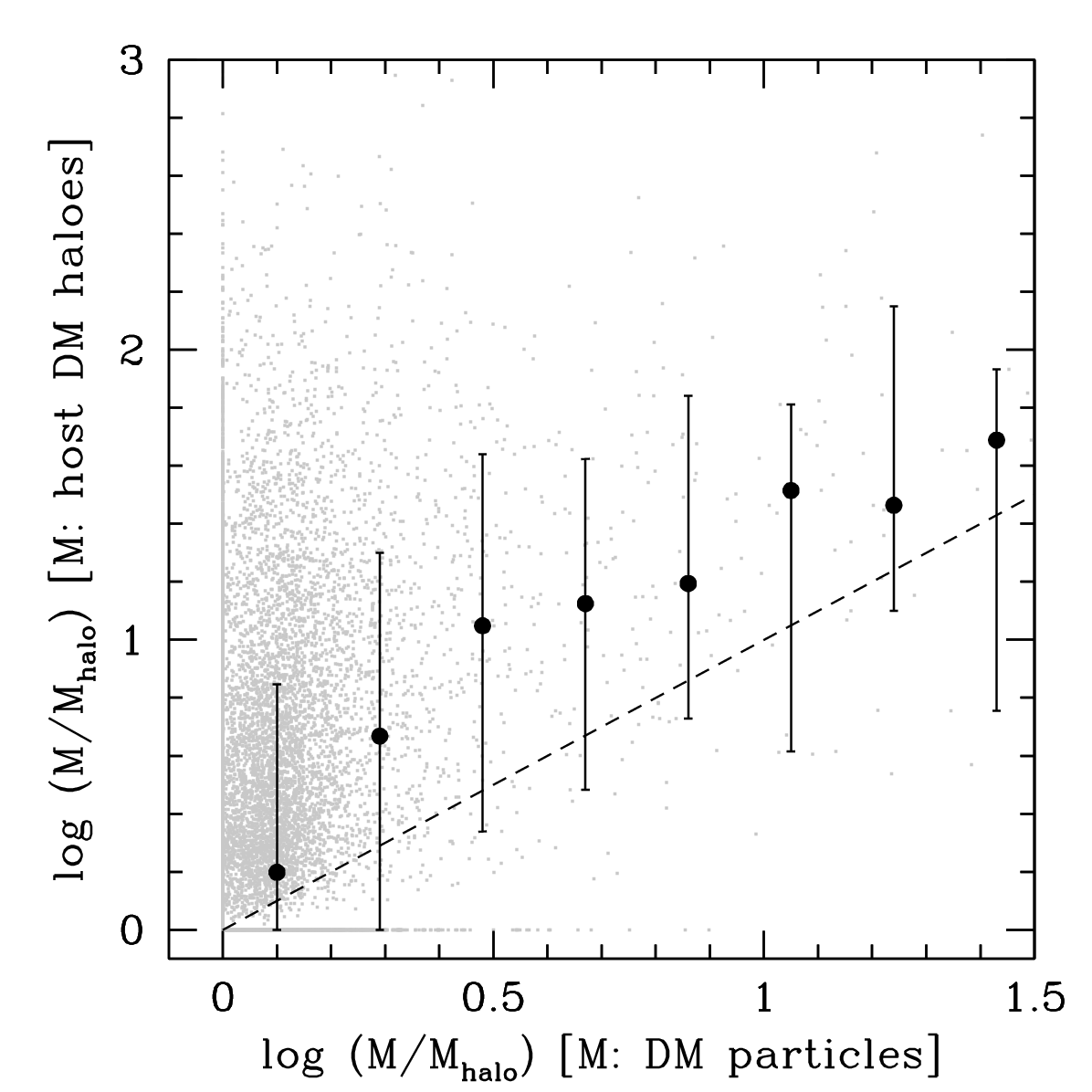}
\caption{
Estimates of the overdensity peak height 
proxy in Section \ref{sec_comp_UnderDis}
using the approach %followed in Paper I 
that considers the mass of DM particles (best-fitting set of parameters $a=0.3$ and $b=-0.12$, x-axis), and the mass of  
host DM haloes 
($a=0.1$ and $b=-0.84$, y-axis).
This is the total mass contained in spheres and cylinders, respectively, around the same synthetic central
galaxies (grey dotted points) in units of the virial mass of their host DM haloes. 
%are the result for a random sample of 5,000
Solid circles are the medians and the error bars correspond to the 16 and 84 percentiles of the distribution. 
%Solid lines are isocontours of number density increasing by 0.5 dex
%(there are very few galaxies at log($M/M_{halo}$) $\geq 1.5$ at the x-axis).}
The dashed line represents a one-to-one relation for both masses.
}
\label{comp_mass}
\end{center}
\end{figure}

The discrete formalism presented in Section \ref{sec_prm}
actually measures the crowding
around objects
which constitutes a different 
proxy of the peak height 
than the total underlying mass.
In order to compare both proxies of the peak height, we use the semi-analytical galaxies that were used to construct
the mock catalogue (see Section \ref{section_sam}). The discrete mass proxy is performed by means of the total mass of neighbour host haloes within cylinders around synthetic central galaxies with the condition 
$M_r -$ 5 log$(h) \le -19.6$.
Only neighbour host haloes with masses $> 10^{11.6} h^{-1}$ $M_{\odot}$ are considered.
On the other hand, the underlying mass proxy is estimated by using the DM particles of the numerical simulation (Section \ref{section_sam}) contained in spheres around the same synthetic central galaxies. In both cases, the radius of the cylinders/spheres is given by the equation (\ref{eq_r_prm}). 
The length of each cylinder is $\Delta v$ = $\pm$ 500 $\kms$ centred at synthetic central galaxies by using 
the $z$-axis of the simulation as the
line-of-sight and turn the $z$-coordinate into a recession
velocity, $v'_z = 100 z + v_z$, where $v_z$ is the velocity component in
the $z$-coordinate. 
We then repeat the procedure described in Section \ref{sec_SAMgalx}, but using cross-correlation functions in real space in equation (\ref{eq_chixi}).
The best-fitting parameters obtained for the discrete mass proxy are
$a = 0.1, b= -0.84$, which are consistent with the marginalized likelihood functions of the mock catalogue (Fig. \ref{likelihoods}). %using similar discrete tracers.
The best-fitting values for the underlying mass proxy are $a = 0.3, b= -0.12$.
Fig. \ref{comp_mass} shows a comparison between the two estimates of the overdensity peak height proxy. The approach that considers the mass of host DM haloes adds on average more mass than the approach using the mass of DM particles, which results in a different mass function (c.f. Fig. \ref{MF}).
However,
%that
%The former measures higher masses than the approach
%followed in Paper I because the radii are typically greater,
%but 
there is a clear correlation between both methods,
%showing
indicating that the same phenomenon is being 
described. %to a high degree.
We calculate a 
correlation coefficient using the Pearson's product-moment coefficient,
\begin{eqnarray}
C = \frac{\displaystyle \sum_{i=1}^N (\Delta d_i - \overline{\Delta d}) (\Delta u_i - \overline{\Delta u})}
{\displaystyle \sqrt{ \sum_{i=1}^N (\Delta d_i - \overline{\Delta d})^2 } \sqrt{\sum_{i=1}^N (\Delta u_i - \overline{\Delta u})^2 }} \textrm{ ,}
\end{eqnarray}
%\\
where $\Delta d$ = log($M/M_{h}$) for the discrete mass proxy (y-axis of Fig. \ref{comp_mass}) and $\Delta u$ = log($M/M_{h}$) for the underlying mass proxy (x-axis of Fig. \ref{comp_mass}).
Their respective averages are $\overline{\Delta d}$ and $\overline{\Delta u}$.
We obtain %$C = 0.41$, 
$C = 0.36$, which means that there is 
a medium strength in the correlation
between both proxies of the peak height, at least with a lower halo mass limit $M_{h} = 10^{11.6} h^{-1}$ $M_{\odot}$.
This strength in the correlation does not change if we include in the discrete proxy a scatter in the halo mass of the simulation similar to that found by Y07.

We conclude that although our discrete proxy to measure the total mass %does
will not reproduce %a reliable distribution of luminosity,
the mass function, %of haloes in the simulation, 
the reason for this is only the use of discrete %tracers.
neighbours.
If we were able to measure the %total light 
distribution of mass around central galaxies, 
we would obtain a more physical mass function (e.g. Paper I).
This result implies that the phenomena behind the assembly bias,
namely, the truncation of infall mass in haloes of low mass embedded in high density regions, can be detected with this discrete proxy for the peak height that can be applied to real galaxy surveys.

%%%section
\section{Conclusions}
\label{conclusiones}

We have shown that the assembly bias effect, namely, the difference in the clustering
amplitude at large scales of populations of equal mass but different age, exists for both mock central galaxies and
SDSS central galaxies. 
This is a relevant issue that could affect 
the ability of
the next generation 
of galaxy surveys 
to infer accurate cosmological parameters.
By means of a  
large halo-based galaxy group catalogue (Y07), constructed from the SDSS, we select central galaxies (i.e. the most massive galaxy in the group). This catalogue includes the estimation of the host halo mass with a lower limit of $M_{h} = 10^{11.6} h^{-1}$ $M_{\odot}$.
Using the
projected correlation function, 
%old galaxies have systematically higher clustering amplitude than
%young ones of the same magnitude at scales $r > 1$ $h^{-1}$ Mpc.
the clustering strength of old (mock) SDSS central galaxies
is (35$-$55 per cent) 50$-$70 per cent higher than that of
young (mock) SDSS central galaxies of same halo mass ($M_{h} \sim 10^{11.8} h^{-1}$ $M_{\odot}$) at projected scales $ > 1$ $h^{-1}$ Mpc. 
%where
%the simulated galaxies show a more important effect.
The estimator of age is based on the mass- and luminosity-weighted stellar ages for the synthetic and SDSS galaxies, respectively.

%Lacerna $\&$ Padilla (2011, Paper I) 
\citet[][Paper I]{Lacerna11} presented an overdensity peak height proxy with the aim to understand the assembly bias effect. This new definition was proposed as a better alternative than the virial mass. 
%for which the large-scale clustering
%of objects of equal mass did not depend on the age. 
In this work we adapt this model
to observations.
%using galaxy luminosities,
%taking advantage of their correlation 
%with substructure mass.
As the SDSS catalogue is limited in flux, and as the tracers of mass are the host haloes of neighbours central galaxies, the method is modified to use the crowding of the environment around central galaxies, %in selected volume-limited samples, 
instead of searching for a correction to the virial mass as was done in Paper I.
We measure the total mass given by the mass of neighbour host haloes in cylinders centred at each central galaxy, thus %measuring the density
obtaining an estimate of the crowding around them, which traces the assembly bias. The radius of this cylinder is parametrized as a function of stellar age and host halo mass of the central galaxy. The best-fitting sets of parameters in the mock catalogue and observations are similar
($a=0.3$, $b=-0.6$ and $a=0.3$, $b=-1.08$, respectively).
In both cases
the assembly bias is %almost not present
lower than 5$-$15 per cent
after using this approach.
%The average differences in clustering amplitude
%between old and young populations for both synthetic and
%observational galaxies are smaller than a 10 percent
%for each range in magnitude $M'_r$.
By estimating a weighted average for the values in the range of $1\sigma$ for mock and SDSS central galaxies,
it is obtained that the set of best-fitting parameters for the radius of the cylinder is
$a = 0.26 \pm 0.19$,
$b = -0.75 \pm 0.36$.

This latter model, which %resembles
constitutes a discrete formalism when counting
neighbour haloes within cylinders, %(or spheres), 
does not reproduce
the original
mass functions %though, but
since it  actually measures the crowding around central galaxies.
%although
However, it is %highly 
reasonably correlated with the smooth density field approach of Paper I.
Therefore, both formalisms %are associated with the fact that
help us reach the same conclusion, that
the virial mass of %peaks that did not collapse on to haloes will not be an appropriate 
dark matter halos sometimes is not appropriate as a
proxy for the peak height.
It is likely that 
low-mass haloes in high density environments show their growth truncated by the gravitational effects of their massive neighbours.
Our results indicate that this phenomenon also affects the 
growth of %the stellar mass in 
galaxies in addition to 
the growth of their host dark matter haloes.
%, however, given by the higher
%differences in the clustering amplitude of these objects.
This might induce a misclassification
of galaxies in theoretical models to populate haloes,
such as the Halo Occupation Distribution (HOD), because of using virial mass instead of proper peak height.

%This kind of studies
These results will be of high importance 
for the next
generation of galaxy surveys, such as LSST. The huge amount of information
available will allow us to measure the clustering amplitude of galaxies with
significant statistics at Mpc scales. Therefore, as we show in this paper, it will be absolutely necessary to model the effect of the environment on the two-halo regime in order to understand its role in the formation and evolution of haloes and galaxies.
Theoretical models will have to include this phenomenon in order to reproduce the observational data with high accuracy
in the new era of precision cosmology.  
%will shed light
%about the implication of the cosmological models

\section*{Acknowledgments}

%IL thanks 
%travel support to attend 
%international conferences from Fondo ALMA-CONICYT 31070007 and VRAID.
We would like to thank Vladimir Avila-Reese for comments and discussions.
IL acknowledges support from the Postdoctoral Fellowship program of DGAPA-UNAM, Mexico. 
NP acknowledges support from Fondecyt No. 1110328, BASAL PFB-06 ``Centro de Astrofisica y Tecnologias Afines".  
FS is supported by the DFG Research Unit FOR 1254.
Part of these calculations were performed using the Geryon cluster at AIUC, which received joint funding from Anillo ACT-86, FONDEQUIP AIC-57, and QUIMAL 130008.

%The Millennium Simulation databases used in this paper and the web application providing online access to them were constructed as part of the activities of the German Astrophysical Virtual Observatory.

\bibliography{references}

\begin{thebibliography}{45}
\providecommand{\natexlab}[1]{#1}

\bibitem[{{Adelman-McCarthy} et~al.(2006){Adelman-McCarthy}, {Ag{\"u}eros},
  {Allam}, {Anderson} \& {et al.}}]{DR4+2006}
{Adelman-McCarthy} J.~K., {Ag{\"u}eros} M.~A., {Allam} S.~S., {Anderson}
  K.~S.~J., {et al.}, 2006, \apjs, 162, 38

\bibitem[{{Alonso} et~al.(2012){Alonso}, {Mesa}, {Padilla} \&
  {Lambas}}]{Alonso12}
{Alonso} S., {Mesa} V., {Padilla} N., {Lambas} D.~G., 2012, \aap, 539, A46

\bibitem[{{Berlind} \& {Weinberg}(2002)}]{BW02}
{Berlind} A.~A., {Weinberg} D.~H., 2002, \apj, 575, 587

\bibitem[{{Berrier} et~al.(2011){Berrier}, {Barton}, {Berrier}, {Bullock},
  {Zentner} \& {Wechsler}}]{Berrier+11}
{Berrier} H.~D., {Barton} E.~J., {Berrier} J.~C., {Bullock} J.~S., {Zentner}
  A.~R., {Wechsler} R.~H., 2011, \apj, 726, 1

\bibitem[{{Blanton} et~al.(2005)}]{Blanton+2005}
{Blanton} M.~R. et~al., 2005, \aj, 129, 2562

\bibitem[{{Cabr{\'e}} et~al.(2007){Cabr{\'e}}, {Fosalba}, {Gazta{\~n}aga} \&
  {Manera}}]{Cabre07}
{Cabr{\'e}} A., {Fosalba} P., {Gazta{\~n}aga} E., {Manera} M., 2007, \mnras,
  381, 1347

\bibitem[{{Colless} et~al.(2001)}]{Colless01}
{Colless} M., {Dalton} G., {Maddox} S., {Sutherland} W., {Norberg} P., {Cole}
  S., {Bland-Hawthorn} J., {et al.}, 2001, \mnras, 328, 1039

\bibitem[{{Cooper} et~al.(2010){Cooper}, {Gallazzi}, {Newman} \&
  {Yan}}]{Cooper10}
{Cooper} M.~C., {Gallazzi} A., {Newman} J.~A., {Yan} R., 2010, \mnras, 402,
  1942

\bibitem[{{Cora}(2006)}]{Cora06}
{Cora} S.~A., 2006, \mnras, 368, 1540

\bibitem[{{Croton} et~al.(2007){Croton}, {Gao} \& {White}}]{Croton07}
{Croton} D.~J., {Gao} L., {White} S.~D.~M., 2007, \mnras, 374, 1303

\bibitem[{{Dalal} et~al.(2008){Dalal}, {White}, {Bond} \& {Shirokov}}]{Dalal08}
{Dalal} N., {White} M., {Bond} J.~R., {Shirokov} A., 2008, \apj, 687, 12

\bibitem[{{Davis} et~al.(1985){Davis}, {Efstathiou}, {Frenk} \&
  {White}}]{Davis85}
{Davis} M., {Efstathiou} G., {Frenk} C.~S., {White} S.~D.~M., 1985, \apj, 292,
  371

\bibitem[{{Faltenbacher} \& {White}(2010)}]{FW10}
{Faltenbacher} A., {White} S.~D.~M., 2010, \apj, 708, 469

\bibitem[{{Gallazzi} et~al.(2005){Gallazzi}, {Charlot}, {Brinchmann}, {White}
  \& {Tremonti}}]{Gallazzi05}
{Gallazzi} A., {Charlot} S., {Brinchmann} J., {White} S.~D.~M., {Tremonti}
  C.~A., 2005, \mnras, 362, 41

\bibitem[{{Gao} \& {White}(2007)}]{Gao-White07}
{Gao} L., {White} S.~D.~M., 2007, \mnras, 377, L5

\bibitem[{{Gao} et~al.(2005){Gao}, {Springel} \& {White}}]{Gao05}
{Gao} L., {Springel} V., {White} S.~D.~M., 2005, \mnras, 363, L66

\bibitem[{{Hahn} et~al.(2009){Hahn}, {Porciani}, {Dekel} \& {Carollo}}]{Hahn09}
{Hahn} O., {Porciani} C., {Dekel} A., {Carollo} C.~M., 2009, \mnras, 398, 1742

\bibitem[{{Hearin} et~al.(2014){Hearin}, {Watson} \& {van den
  Bosch}}]{Hearin+2014}
{Hearin} A.~P., {Watson} D.~F., {van den Bosch} F.~C., 2014, ArXiv e-prints

\bibitem[{{Kauffmann} et~al.(2013){Kauffmann}, {Li}, {Zhang} \&
  {Weinmann}}]{Kauffmann+2013}
{Kauffmann} G., {Li} C., {Zhang} W., {Weinmann} S., 2013, \mnras, 430, 1447

\bibitem[{{Lacerna} \& {Padilla}(2011)}]{Lacerna11}
{Lacerna} I., {Padilla} N., 2011, \mnras, 412, 1283

\bibitem[{{Lacerna} \& {Padilla}(2012)}]{Lacerna12}
{Lacerna} I., {Padilla} N., 2012, \mnras, 426, L26

\bibitem[{{Lagos} et~al.(2008){Lagos}, {Cora} \& {Padilla}}]{LCP08}
{Lagos} C.~D.~P., {Cora} S.~A., {Padilla} N.~D., 2008, \mnras, 388, 587

\bibitem[{{Li} et~al.(2013){Li}, {Gao}, {Xie} \& {Guo}}]{Li13}
{Li} R., {Gao} L., {Xie} L., {Guo} Q., 2013, \mnras, 435, 3592

\bibitem[{{Norberg} et~al.(2009){Norberg}, {Baugh}, {Gazta{\~n}aga} \&
  {Croton}}]{Norberg09}
{Norberg} P., {Baugh} C.~M., {Gazta{\~n}aga} E., {Croton} D.~J., 2009, \mnras,
  396, 19

\bibitem[{{Skibba} et~al.(2006){Skibba}, {Sheth}, {Connolly} \&
  {Scranton}}]{Skibba06}
{Skibba} R., {Sheth} R.~K., {Connolly} A.~J., {Scranton} R., 2006, \mnras, 369,
  68

\bibitem[{{Skibba} \& {Sheth}(2009)}]{Skibba_Sheth_2009}
{Skibba} R.~A., {Sheth} R.~K., 2009, \mnras, 392, 1080

\bibitem[{{Springel} et~al.(2001){Springel}, {White}, {Tormen} \&
  {Kauffmann}}]{Springel01}
{Springel} V., {White} S.~D.~M., {Tormen} G., {Kauffmann} G., 2001, \mnras,
  328, 726

\bibitem[{{Tinker} et~al.(2011){Tinker}, {Wetzel} \& {Conroy}}]{Tinker11}
{Tinker} J., {Wetzel} A., {Conroy} C., 2011, ArXiv e-prints

\bibitem[{{van Daalen} et~al.(2012){van Daalen}, {Angulo} \&
  {White}}]{vanDaalen12}
{van Daalen} M.~P., {Angulo} R.~E., {White} S.~D.~M., 2012, \mnras, 424, 2954

\bibitem[{{Wang} et~al.(2007){Wang}, {Mo} \& {Jing}}]{WangH07}
{Wang} H.~Y., {Mo} H.~J., {Jing} Y.~P., 2007, \mnras, 375, 633

\bibitem[{{Wang} et~al.(2013{\natexlab{a}}){Wang}, {De Lucia} \&
  {Weinmann}}]{WangDeLuciaWeinmann13}
{Wang} L., {De Lucia} G., {Weinmann} S.~M., 2013{\natexlab{a}}, \mnras, 431,
  600

\bibitem[{{Wang} et~al.(2013{\natexlab{b}}){Wang}, {Weinmann}, {De Lucia} \&
  {Yang}}]{WangL13}
{Wang} L., {Weinmann} S.~M., {De Lucia} G., {Yang} X., 2013{\natexlab{b}},
  \mnras, 433, 515

\bibitem[{{Wang} et~al.(2008){Wang}, {Yang}, {Mo}, {van den Bosch}, {Weinmann}
  \& {Chu}}]{Wang08}
{Wang} Y., {Yang} X., {Mo} H.~J., {van den Bosch} F.~C., {Weinmann} S.~M.,
  {Chu} Y., 2008, \apj, 687, 919

\bibitem[{{Wechsler} et~al.(2006){Wechsler}, {Zentner}, {Bullock}, {Kravtsov}
  \& {Allgood}}]{Wechsler06}
{Wechsler} R.~H., {Zentner} A.~R., {Bullock} J.~S., {Kravtsov} A.~V., {Allgood}
  B., 2006, \apj, 652, 71

\bibitem[{{Wu} et~al.(2008){Wu}, {Rozo} \& {Wechsler}}]{Wu08}
{Wu} H.~Y., {Rozo} E., {Wechsler} R.~H., 2008, \apj, 688, 729

\bibitem[{{Xie} et~al.(2014){Xie}, {Gao} \& {Guo}}]{Xie14}
{Xie} L., {Gao} L., {Guo} Q., 2014, \mnras, 441, 933

\bibitem[{{Yang} et~al.(2003){Yang}, {Mo} \& {van den Bosch}}]{Y03}
{Yang} X., {Mo} H.~J., {van den Bosch} F.~C., 2003, \mnras, 339, 1057

\bibitem[{{Yang} et~al.(2006){Yang}, {Mo} \& {van den Bosch}}]{Y06}
{Yang} X., {Mo} H.~J., {van den Bosch} F.~C., 2006, \apjl, 638, L55

\bibitem[{{Yang} et~al.(2007){Yang}, {Mo}, {van den Bosch}, {Pasquali}, {Li} \&
  {Barden}}]{Yang+2007}
{Yang} X., {Mo} H.~J., {van den Bosch} F.~C., {Pasquali} A., {Li} C., {Barden}
  M., 2007, \apj, 671, 153

\bibitem[{{Yang} et~al.(2008){Yang}, {Mo} \& {van den Bosch}}]{Yang+2008}
{Yang} X., {Mo} H.~J., {van den Bosch} F.~C., 2008, \apj, 676, 248

\bibitem[{{York} et~al.(2000){York}, {Adelman}, {Anderson}, {Anderson}, {Annis}
  \& {et al.}}]{York00}
{York} D.~G., {Adelman} J., {Anderson} Jr. J.~E., {Anderson} S.~F., {Annis} J.,
  {et al.}, 2000, \aj, 120, 1579

\bibitem[{{Zapata} et~al.(2009){Zapata}, {Perez}, {Padilla} \&
  {Tissera}}]{Zapata09}
{Zapata} T., {Perez} J., {Padilla} N., {Tissera} P., 2009, \mnras, 394, 2229

\bibitem[{{Zehavi} et~al.(2002)}]{Zehavi02}
{Zehavi} I. et~al., 2002, \apj, 571, 172

\bibitem[{{Zentner} et~al.(2013){Zentner}, {Hearin} \& {van den
  Bosch}}]{Zentner13}
{Zentner} A.~R., {Hearin} A.~P., {van den Bosch} F.~C., 2013, ArXiv e-prints

\bibitem[{{Zhu} et~al.(2006){Zhu}, {Zheng}, {Lin}, {Jing}, {Kang} \&
  {Gao}}]{Zhu06}
{Zhu} G., {Zheng} Z., {Lin} W.~P., {Jing} Y.~P., {Kang} X., {Gao} L., 2006,
  \apjl, 639, L5

\end{thebibliography}

\label{lastpage}

\end{document}